\documentclass[a4paper,11pt]{article} 
\pdfoutput=1 
\usepackage{jinstpub} 

\usepackage{caption}
\usepackage{subcaption}
\usepackage{mathrsfs}
\graphicspath{{figures/}{/plots}} 

\usepackage{xspace}

\title{Radiation hardness of small-pitch 3D pixel sensors up to a fluence of $3\times10^{16}$\,n$_{\mathrm{eq}}$/cm$^2$}

\author[a,1]{J.~Lange\note{Corresponding author.},} 
\author[a]{G.~Giannini,}
\author[a,c]{S.~Grinstein,}
\author[a,b]{M.~Manna,}
\author[b]{G.~Pellegrini,}
\author[b]{D.~Quirion,}
\author[a]{S.~Terzo,}
\author[a]{D.~V\'{a}zquez Furelos}

\affiliation[a]{Institut de F\'{i}sica d'Altes Energies (IFAE), The Barcelona Institute of Science and Technology (BIST), Campus UAB, 08193 Bellaterra (Barcelona), Spain}

\affiliation[b]{Centro Nacional de Microelectronica (CNM-IMB-CSIC), Campus UAB, 08193 Bellaterra (Barcelona), Spain}

\affiliation[c]{Instituci\'{o} Catalana de Recerca i Estudis Avan\c{c}ats (ICREA), Pg. Llu\'{i}s Companys 23, 08010 Barcelona, Spain}

\emailAdd{joern.lange@cern.ch} 

\abstract{ 

Small-pitch 3D silicon pixel detectors have been investigated as radiation-hard candidates for the innermost layers of the HL-LHC pixel detector upgrades. Prototype 3D sensors with pixel sizes of 50$\times$50 and 25$\times$100\,$\mu$m$^{2}$ connected to the existing ATLAS FE-I4 readout chip have been produced by CNM Barcelona. Irradiations up to particle fluences of $3\times10^{16}$\,n$_{\mathrm{eq}}$/cm$^2$, beyond the full expected HL-LHC fluences at the end of lifetime, have been carried out at Karlsruhe and CERN. The performance of the 50$\times$50\,$\mu$m$^{2}$ devices has been measured in the laboratory and beam tests at CERN SPS. A high charge collected and a high hit efficiency of 98\% were found up to the highest fluence. The bias voltage to reach the target efficiency of 97\% at perpendicular beam incidence was found to be about 100\,V at $1.4\times10^{16}$\,n$_{\mathrm{eq}}$/cm$^2$ and 150\,V at $2.8\times10^{16}$\,n$_{\mathrm{eq}}$/cm$^2$, significantly lower than for the previous IBL 3D generation with larger inter-electrode distance and than for planar sensors. The power dissipation at -25$^{\circ}$C and $1.4\times10^{16}$\,n$_{\mathrm{eq}}$/cm$^2$ was found to be 13\,mW/cm$^2$.
Hence, 3D pixel detectors demonstrated superior radiation hardness and were chosen as the baseline for the inner layer of the ATLAS HL-LHC pixel detector upgrade.

}


\arxivnumber{1805.10208} 

\begin{document}

\maketitle 
\flushbottom 

\section{Introduction}
\label{sec:intro}

Finely segmented silicon pixel detectors are widely used in particle physics for precise position measurements at high occupancies, providing tracks and vertices of particles. The quest for high collider luminosities and operation close to the beam leads to harsh radiation levels, for which radiation-hard sensor technologies such as 3D detectors~\cite{bib:3D} need to be developed. For example, the currently installed innermost pixel layer of the ATLAS experiment at the Large Hadron Collider (LHC) at CERN (Insertable B-layer, IBL) is qualified to withstand a 1-MeV neutron equivalent particle fluence of $\Phi_{\mathrm{eq}}=5\times10^{15}$\,n$_{\mathrm{eq}}$/cm$^2$~\cite{bib:IBL}. For the LHC High-Luminosity upgrade (HL-LHC)~\cite{bib:HL-LHC}, targeting a 10-fold increase of integrated luminosity to 4,000\,fb$^{-1}$, even higher fluences of 2.6$\times10^{16}$\,n$_{\mathrm{eq}}$/cm$^2$ (including a safety factor of 1.5) are expected for the innermost pixel layer of ATLAS at the end of lifetime~\cite{bib:ITkPixelTDR,bib:HL-LHCfluences}. Similar levels are expected for the CMS tracker~\cite{bib:CMSPhase2TrackerTDR}. Hence, replacements of the full tracker systems of both experiments are foreseen for 2024 (called "Inner Tracker" (ITk) in the case of ATLAS). It is furthermore planned to replace the inner layers again after half of the HL-LHC lifetime, reducing the ITk requirement to 1.3$\times10^{16}$\,n$_{\mathrm{eq}}$/cm$^2$ (mainly due to expected limited radiation hardness of the readout electronics)~\cite{bib:ITkPixelTDR}. For potential future collider applications such as the LHCb phase II upgrade~\cite{bib:LHCbPhaseII} or a future hadron circular collider (FCC-hh)~\cite{bib:FCCfluence}, the radiation levels will be even higher with about 1~and 8$\times10^{17}$\,n$_{\mathrm{eq}}$/cm$^2$, respectively.

\begin{figure}[hbtp]
	\centering
	\includegraphics[width=14cm]{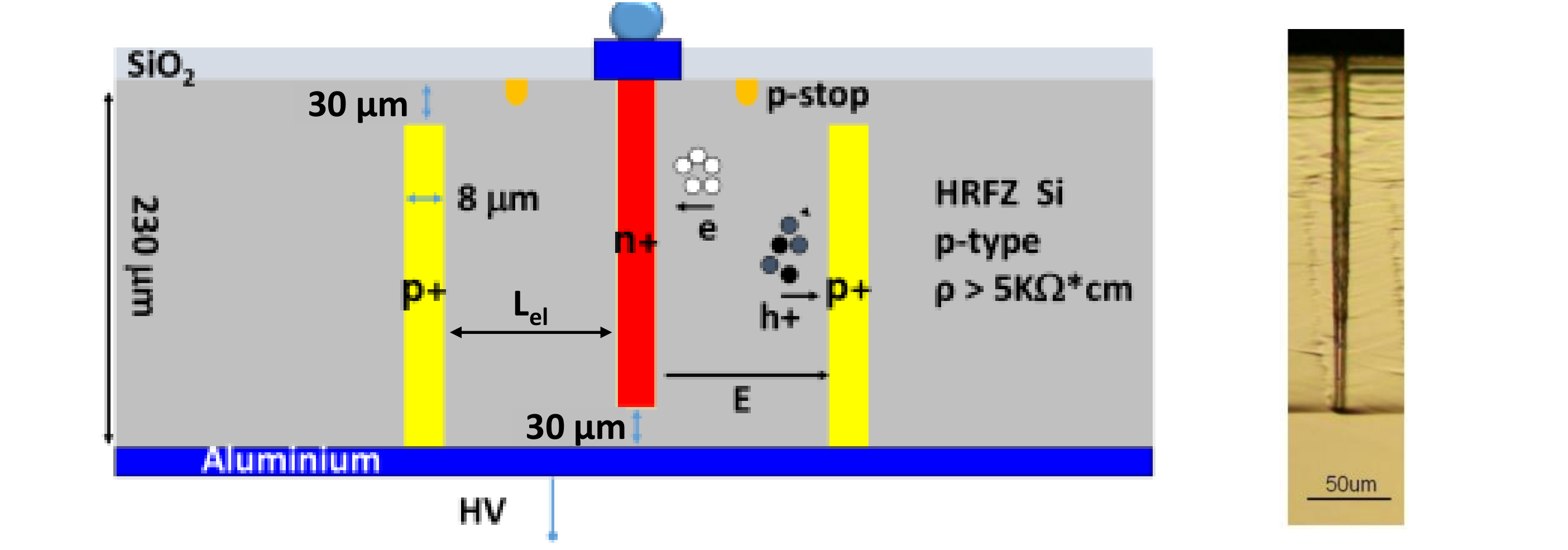}\\
	\vspace{1cm}
	\includegraphics[width=15cm]{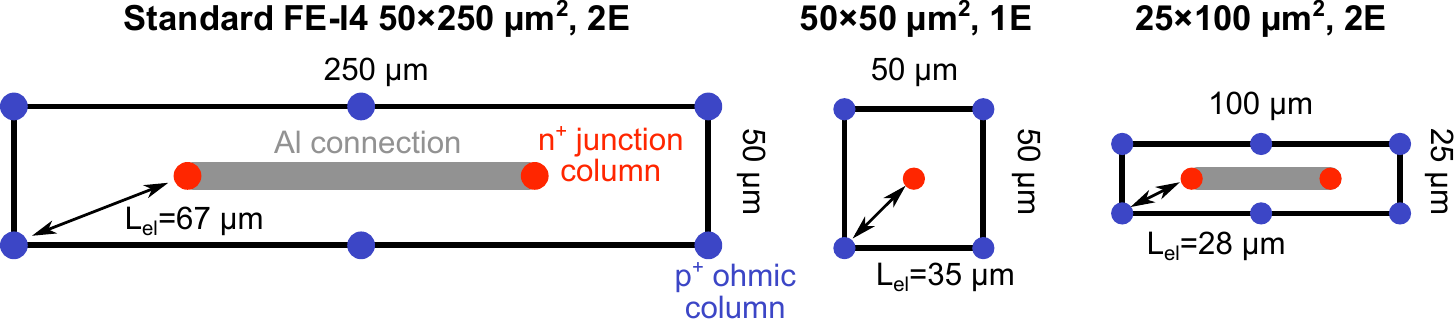}\\
	\vspace{1cm}
	\includegraphics[width=14cm]{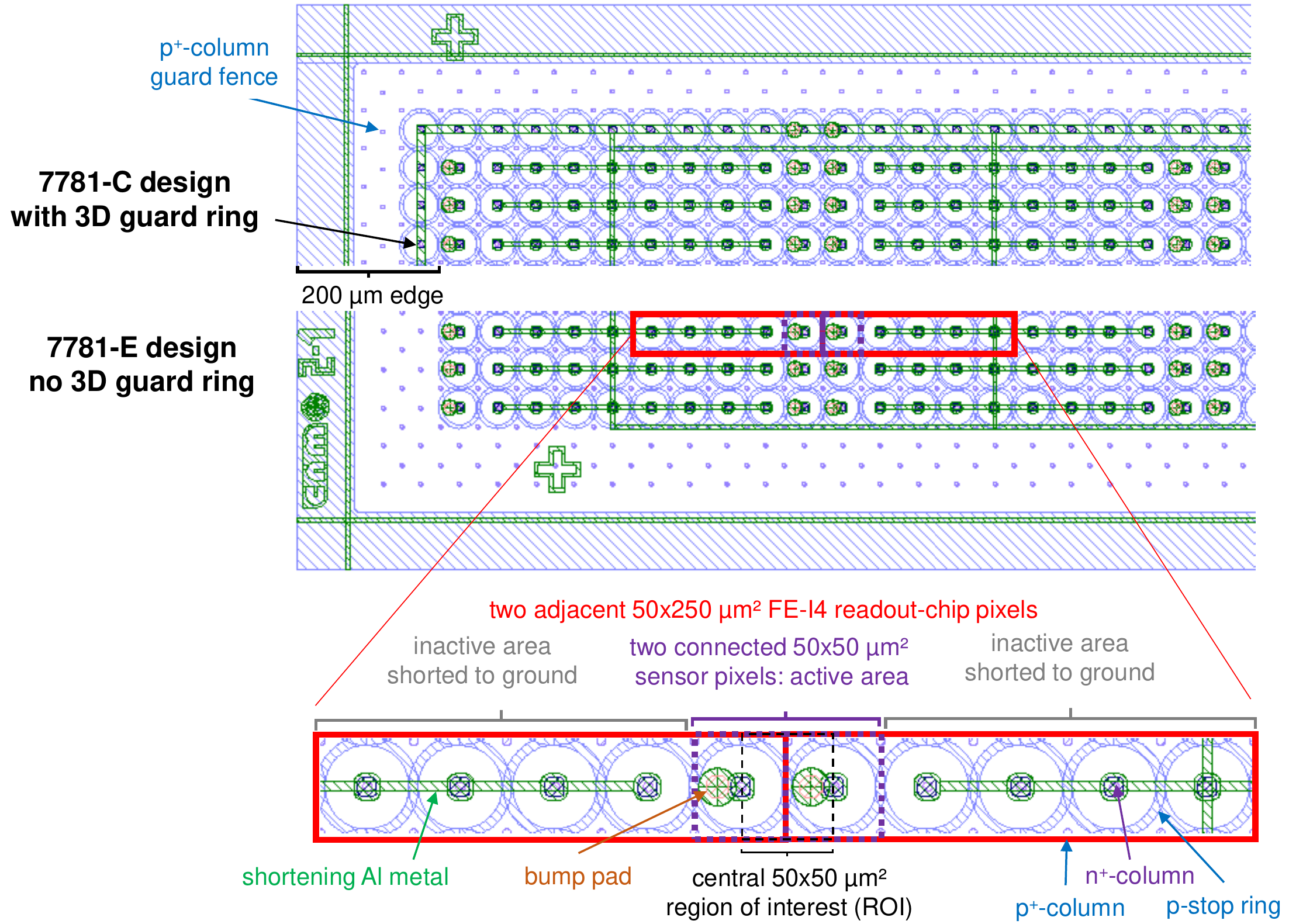}
	\caption{Top left: cross section of a CNM 3D sensor in double-sided technology with non-fully passing through columns. Top right: microscopic picture of a 3D column of run 7781. Centre: geometry of a 3D pixel cell for a standard IBL/AFP FE-I4 pixel with 50$\times$250\,$\mu$m$^{2}$ 2E configuration (left), and for a 50$\times$50\,$\mu$m$^{2}$ 1E, (centre) and a 25$\times$100\,$\mu$m$^{2}$ 2E, (right) pixel. Bottom: layout of a small-pitch 3D pixel prototype sensor of run 7781 matched to the FE-I4 readout chip in two different edge designs: with ("C") and without ("E") 3D guard ring.}
	\label{fig:3Dgeometry}
\end{figure}

At such high fluences, the major limitation due to radiation damage is the trapping of signal charge carriers at radiation-induced defects along their drift to the collecting electrodes. 3D silicon detectors represent a radiation-hard sensor technology, since the electrodes penetrate the sensor bulk as columns perpendicular to the surface instead of being implanted at the top and bottom surfaces as for planar detectors (see figure~\ref{fig:3Dgeometry}). Hence, the inter-electrode distance L$_{\mathrm{el}}$ is decoupled from the device thickness, allowing to keep the sensor thick enough for a large signal charge deposited by an ionising particle, while reducing L$_{\mathrm{el}}$ to a level where trapping at defects is less severe~\cite{bib:geometryDependence}. This leads to larger charge collection with lower operational voltages and power dissipation compared to planar detectors after irradiation. During the last years, the 3D technology has matured, and by now 3D pixel detectors are already used in high-energy physics particle detectors where superior radiation hardness is key: in the ATLAS IBL~\cite{bib:IBL,bib:IBLprototypes,bib:IBL3Dprod}, the ATLAS Forward Proton (AFP) detector~\cite{bib:AFPTDR,bib:AFP3D2, bib:AFPbeamTests, bib:AFPproduction} and the CMS-TOTEM Precision Proton Spectrometer~\cite{bib:PPS}. 

This first generation of 3D pixel detectors applied in experiments consists e.g. in ATLAS of 50$\times$250\,$\mu$m$^{2}$ pixel cells with two 3D readout electrodes each (2E), which are connected to the FE-I4 readout chip~\cite{bib:IBLprototypes,bib:FEI4} (see figure~\ref{fig:3Dgeometry} centre left). Their 3D inter-electrode distance is L$_{\mathrm{el}}$=67\,$\mu$m, while their sensor thickness is 230\,$\mu$m. This generation demonstrated already a good radiation hardness up to at least $9\times10^{15}$\,n$_{\mathrm{eq}}$/cm$^2$~\cite{bib:CNMIBLgenAndSmallPitch2}. 

For the HL-LHC, pixel detectors with 50$\times$50 or 25$\times$100\,$\mu$m$^{2}$ pixel size are planned, primarily motivated by particle occupancy. Consequently, a new generation of small-pitch 3D detectors is being developed, which also leads to a significant reduction of L$_{\mathrm{el}}$ down to e.g. 35\,$\mu$m for 50$\times$50\,$\mu$m$^{2}$ pixel size with one 3D electrode per pixel (1E) or 27\,$\mu$m for 25$\times$100\,$\mu$m$^{2}$ 2E (see figure~\ref{fig:3Dgeometry} centre). Hence, reduced trapping and improved radiation hardness is expected for a constant thickness. 
Also a new readout chip with 50$\times$50\,$\mu$m$^{2}$ pixel size and improved radiation hardness is under development by the CERN RD53 collaboration~\cite{bib:RD53A}. However, at the time of this study, it has not yet been available. Hence, first small-pitch 3D sensor prototypes have been produced with the option to connect them to the existing FE-I4 and ROC4SENS~\cite{bib:SantanderRoc4Sense} readout chips.

In this work, the performance and radiation hardness of the small-pitch 3D FE-I4 pixel prototypes with 230\,$\mu$m thickness fabricated by CNM (Centro Nacional de Microelectronica) Barcelona has been studied. Emphasis has been laid on the ATLAS ITk baseline fluence range of about 1$\times10^{16}$\,n$_{\mathrm{eq}}$/cm$^2$ (assuming one replacement of the inner layer). First results up to this fluence have been already presented in references~\cite{bib:CNMIBLgenAndSmallPitch2,bib:CNMsmallPitch1,bib:TIPP2017}. Moreover, for the first time, pixel detectors were also irradiated and studied in the laboratory and beam tests up to so far unprecedented fluences of 3$\times10^{16}$\,n$_{\mathrm{eq}}$/cm$^2$, to investigate if this technology could in principle also withstand the full fluence range at the HL-LHC and to perform exploratory measurements towards even higher radiation levels.

\section{Devices}
\label{sec:Devices}

At CNM Barcelona, a 3D sensor run with small pixel sizes of 50$\times$50 and 25$\times$100\,$\mu$m$^{2}$ was carried out as a CERN RD50~\cite{bib:RD50} project (run 7781). The baseline technology and wafers have been the same as for the IBL and AFP productions: double-sided processing of p-type wafers with 230\,$\mu$m thickness and about 20\,k$\Omega$cm resistivity~\cite{bib:CNMIBLProduction}. The n-type junction columns have been etched and doped from the front-side and the p-type ohmic columns from the back side, both stopping about 30\,$\mu$m before reaching the other surface ("non-fully passing through", see figure~\ref{fig:3Dgeometry} top left). The nominal column diameter is 8\,$\mu$m, while the measured values range from 10\,$\mu$m close to the surface, narrowing down to about 5\,$\mu$m near the column tip, as can be seen from figure~\ref{fig:3Dgeometry} top right. Hence, with respect to the 3D CNM IBL production, an average diameter reduction of about 2\,$\mu$m was achieved. Since the RD53A readout chip was still under development during this study, the small 3D sensor pixels are matched to the existing FE-I4 readout chip with larger pixels of 50$\times$250\,$\mu$m$^{2}$ size arranged in a matrix of 336$\times$80 rows and columns: each FE-I4 chip pixel cell contains five sensor pixels of 50$\times$50\,$\mu$m$^{2}$ 1E (labelled "C" or "E" devices) or 25$\times$100\,$\mu$m$^{2}$ 2E geometry (labelled "D"). This layout is shown in figure~\ref{fig:3Dgeometry} bottom for 50$\times$50\,$\mu$m$^{2}$ 1E. Hence, only 20\% of the sensor pixels are connected to a front-end channel and can be read-out. The remaining 80\% insensitive sensor pixels are shorted to ground to be at the same potential as the ones being read-out. The edge termination was either performed with a 3D guard ring of shortened n-type columns around the active area, in combination with a 3D guard fence of p-type columns ("C" and "D" devices), or with a 3D guard fence only ("E") as shown in figure~\ref{fig:3Dgeometry} bottom. The latter design allows the collection of charge beyond the physical boundaries of the last pixel, as seen before in reference~\cite{bib:AFP3D2} and studied for this production in section~\ref{sec:SlimEdge}. The distance between the border of the last pixel and the dicing edge is 200\,$\mu$m.
Also 3D strips and pad diodes of different geometries, as well as other pixel types, were produced. 
Due to the prototyping phase and non-optimised fabrication process parameters, the yield was low (below 50\%), resulting in only about 10 usable small-pitch FE-I4 pixel sensors. However, the yield has been improved in recent productions by tuning the process and protecting the edge of the wafers, resulting e.g. in a wafer (sensor) yield of 83\% (85\%) for the second AFP production~\cite{bib:AFPproduction}. 

On the pixel devices, electro-plate copper or electro-less nickel-gold under-bump metallisation (UBM) was performed at CNM. Then they were bump-bonded to the FE-I4 readout chip at IFAE Barcelona, assembled on readout boards and characterised in the laboratory. The breakdown voltages before irradiation were found to be typically between 15 and 40\,V. The FE-I4 devices were tuned and charge collection studies with a $^{90}$Sr source were performed. Several of them showed areas of disconnected bumps due to poor UBM, whereas others were almost fully connected. Most of the devices were suitable for testing. 

More details of the run and initial laboratory characterisations are described in reference~\cite{bib:CNMsmallPitch1}.


\section{Irradiations}
\label{sec:Irradiations}

On the small-pitch pixel devices, irradiation campaigns were carried out at the Karlsruhe Institute of Technology (KIT)~\cite{bib:KIT} and the CERN-PS IRRAD Proton Facility~\cite{bib:IRRAD,bib:IRRAD2}. KIT delivers 23\,MeV protons with a hardness factor (i.e. ratio of 1-MeV neutron equivalent fluence to proton fluence) of $\kappa=1.85$ and with a Total Ionising Dose (TID) per fluence of 1.5\,MGy per $10^{15}$\,n$_{\mathrm{eq}}$/cm$^2$. In contrast, CERN-PS protons have an energy of 23\,GeV, a hardness factor of $\kappa=0.60$ and a TID per fluence of only 0.4\,MGy per $10^{15}$\,n$_{\mathrm{eq}}$/cm$^2$ due to the higher energy. Details of the devices and irradiations are included in table~\ref{tab:samplesIrradiations}. Several devices have been irradiated multiple times and measured in beam tests after each fluence step.

\begin{table}
	\centering
	\footnotesize
\begin{tabular}{|c|c|c|c|c|c|c|c|c|}																						
\hline											
\multicolumn{2}{|c|}{Device}	&	Irradiation	&	\multicolumn{2}{|c|}{$\Phi_{\mathrm{eq}}$ [$10^{15}$\,n$_{\mathrm{eq}}$/cm$^2$]} &	\multicolumn{2}{|c|}{TID [MGy]}	&	Annealing	&	Status	\\
Name & Geometry      	&	Step+Label	&	Step 	&	Total &	Step 	&	Total	&	Step	&		\\
\hline
\multicolumn{9}{|c|}{CNM Small-Pitch Run 7781, 2016+2017 Irradiations and Beam Tests} \\
\hline											
W3-E& 50x50 1E	&	Unirrad.	&	 - 	&	 - &	 - 	&	 - 	&	 - 	&	Measured	\\
\hline											
W4-D& 25x100 2E	&	PS1	&	15	&	15	& 6.6	&	6.6 &	7d@RT	&	Not working	\\
\hline											
W8-C1& 50x50 1E	&	PS1	&	15	&	15	& 6.6	&	6.6 &	7d@RT	&	Not working	\\
\hline											
W4-C1& 50x50 1E	&	PS1	&	15	&	15	& 6.6	&	6.6 &	7d@RT	&	Measured	\\
	&	&	PS3 &	11	&	26 & 4.8	&	11.4 &	18d@RT	&	Measured	\\
	&	&	PS4 &	6	&	31 & 2.6	&	13.6 &	15d@RT	&	Not working	\\
\hline											
W5-C2& 50x50 1E	&	KIT1 &	4.6	&	4.6 & 6.9	&	6.9 &	8d@RT	&	Measured	\\
	&	&	PS3 &	10	&	15 & 4.4	&	11.3 &	18d@RT	&	Not working	\\
\hline											
W3-C1& 50x50 1E	&	KIT1 &	5.4	&	5.4 & 8.1	&	8.1 &	8d@RT	&	Measured	\\
	&	&	PS2 &	7	&	12 & 3.1	&	11.2 &	15d@RT	&	Working	\\
	&	&	PS3 &	11	&	23 & 4.8	&	16.0 &	18d@RT	&	Measured	\\
	&	&	PS4 &	5	&	28 & 2.2	&	18.2 &	15d@RT	&	Measured	\\
	&	&	PS5 &	3	&	31 & 1.3	&	19.5 &	21d@RT	&	Working	\\
\hline											
W4-E& 50x50 1E	&	KIT2 NoAnn &	10.4	&	10.4 & 15.6	&	15.6 &	0     	&	Measured	\\
	&          	&	KIT2     &	 -  	&	10.4 & -	&	15.6 &	7d@RT	&	Measured	\\
\hline			
\multicolumn{9}{|c|}{CNM IBL Runs (earlier studies~\cite{bib:IBLprototypes,bib:CNMIBLgenAndSmallPitch2})} \\
\hline									
CNM34    & 50x250 2E	&	KIT0	&	 5 	&	 5 & 7.5	&	7.5 &	120min@60$^{\circ}$C 	&	Measured	\\
CNM-NU-1 & 50x250 2E	&	PS0	  &	 6 	&	 6 & 2.6	&	2.6 &	7d@RT 	&	Measured	\\
CNM-NU-2 & 50x250 2E	&	PS0	  &	 9 	&	 9 & 4.0	&	4.0 &	7d@RT 	&	Measured	\\
\hline
							
\end{tabular}											

	\caption{Samples and irradiations. For non-uniform irradiation at PS, the fluence and TID values refer to the peak.}
	\label{tab:samplesIrradiations}
\end{table}

At KIT, two 50$\times$50\,$\mu$m$^{2}$ pixel devices (W3-C1 and W5-C2) have been irradiated uniformly with 23\,MeV protons to 5$\times10^{15}$\,n$_{\mathrm{eq}}$/cm$^2$ and one device (W4-E) to 1.0$\times10^{16}$\,n$_{\mathrm{eq}}$/cm$^2$. The devices were irradiated at a temperature of -36$^{\circ}$C and annealed for about one week at room temperature ($\approx22^{\circ}$C) after irradiation, referred to as "1week@RT" in the following. The dosimetry has been performed using activation of a nickel foil with an accuracy of 10\%. This irradiation type and the lower fluence step of 5$\times10^{15}$\,n$_{\mathrm{eq}}$/cm$^2$ is the same as used in the IBL qualification campaigns~\cite{bib:IBLprototypes} and hence allows a direct comparison to the IBL generation. The higher fluence step is close to the ITk baseline target fluence, assuming one replacement of the inner layer. The fluence uniformity allows a simple analysis as well as determination of the power dissipation. However, the low-energy protons deliver a high TID to the FE-I4 readout chip and hence higher fluences are likely to damage the chip.

\begin{figure}[hbtp]
	\centering
	\includegraphics[width=7cm]{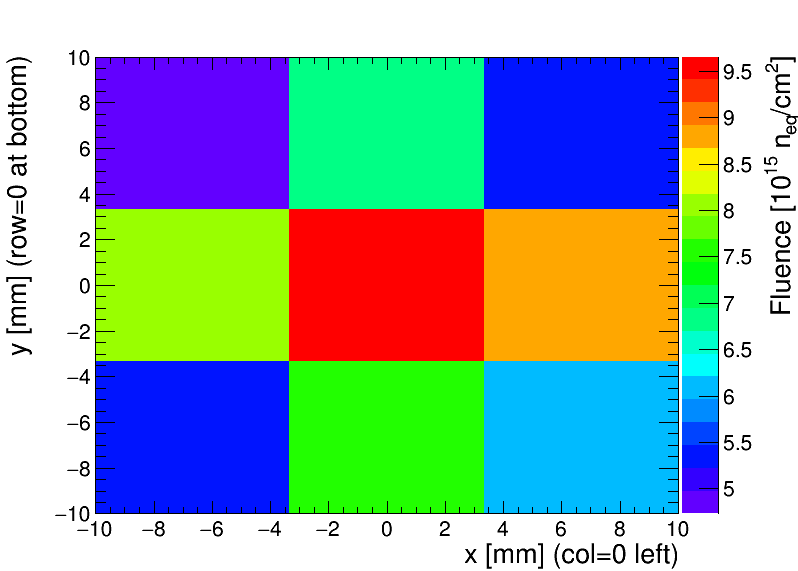}
	\includegraphics[width=7cm]{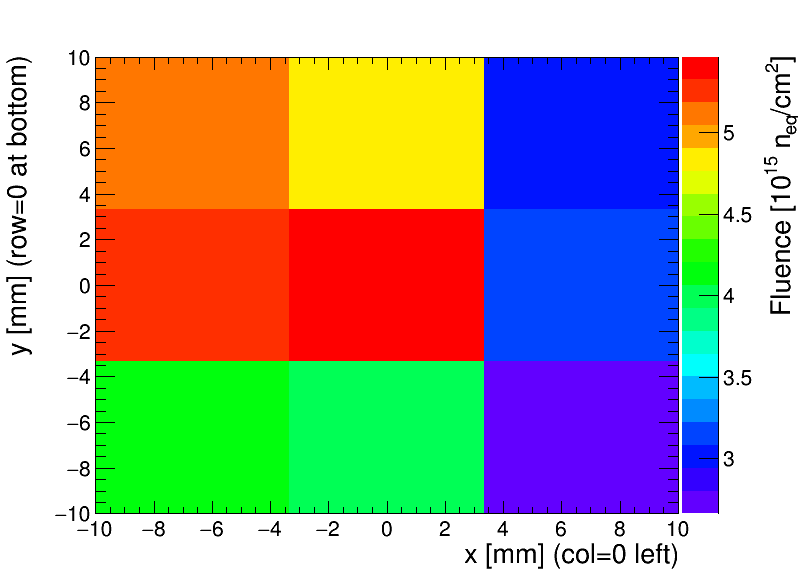}
	\caption{Measured beam profiles of the PS3 (left) and PS4 (right) beam periods using a matrix of Al sub-foils.}
	\label{fig:fluenceMapsAl}
\end{figure}

\begin{figure}[hbtp]
	\centering
	\includegraphics[width=3.7cm]{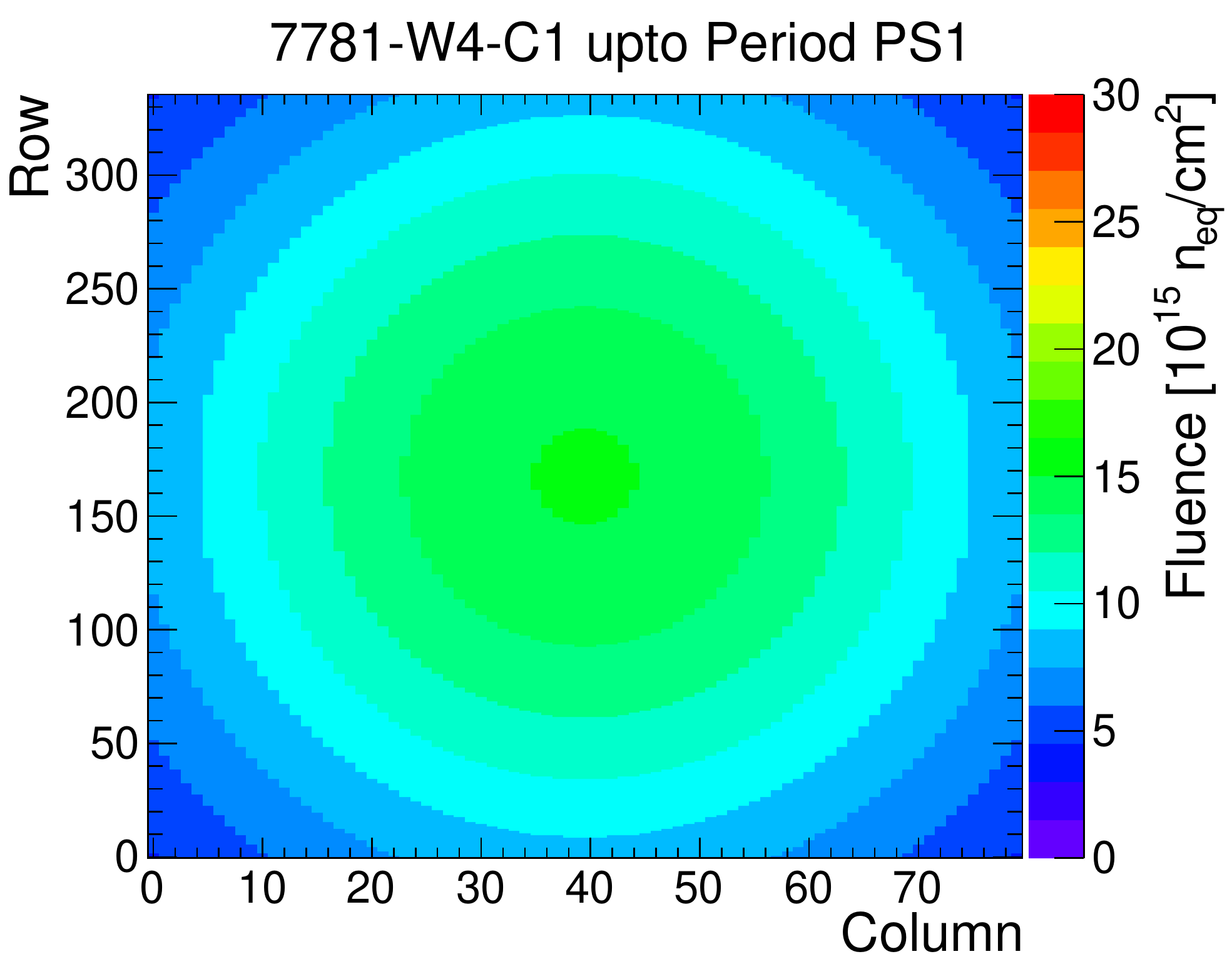}
	\includegraphics[width=3.7cm]{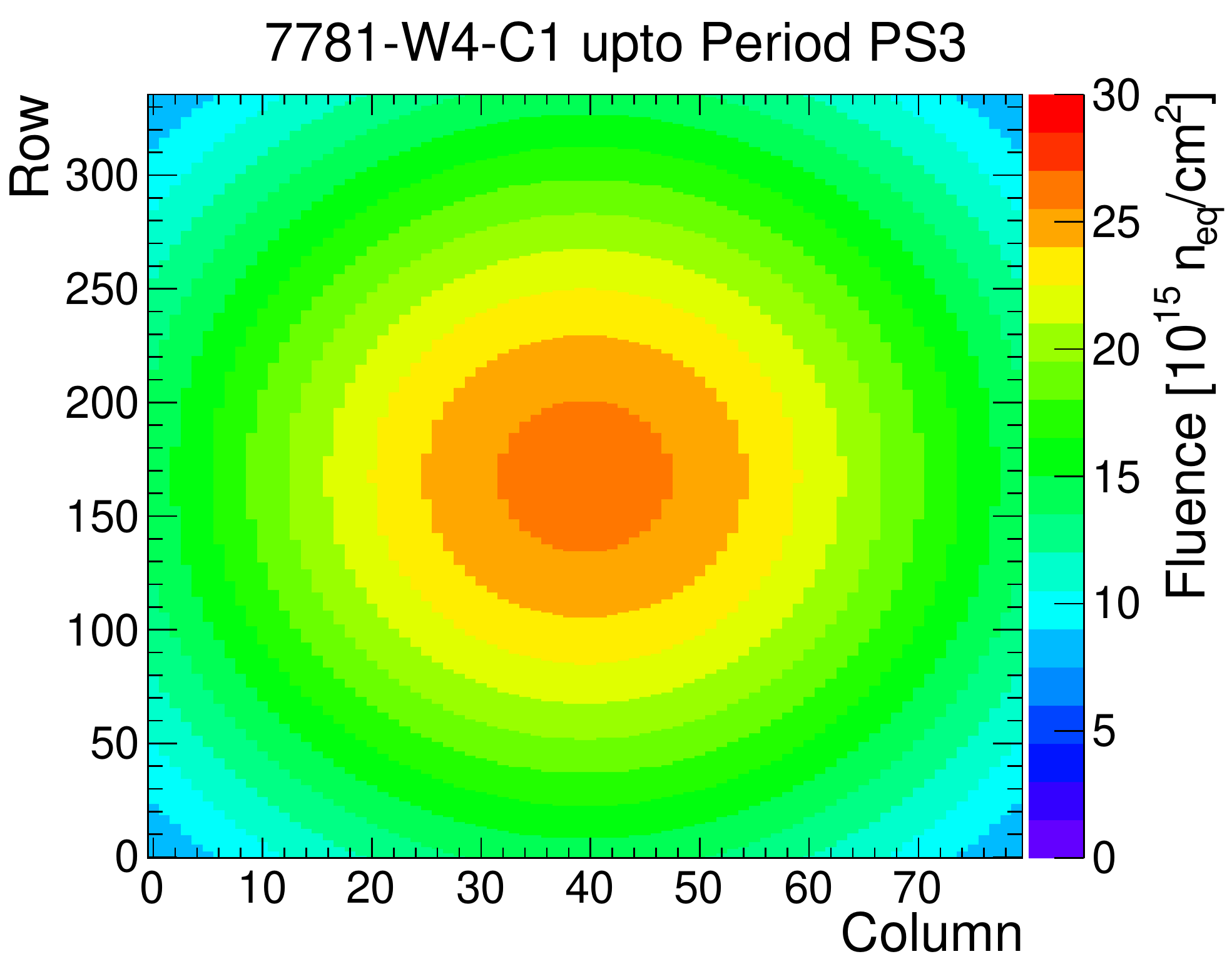}
	\includegraphics[width=3.7cm]{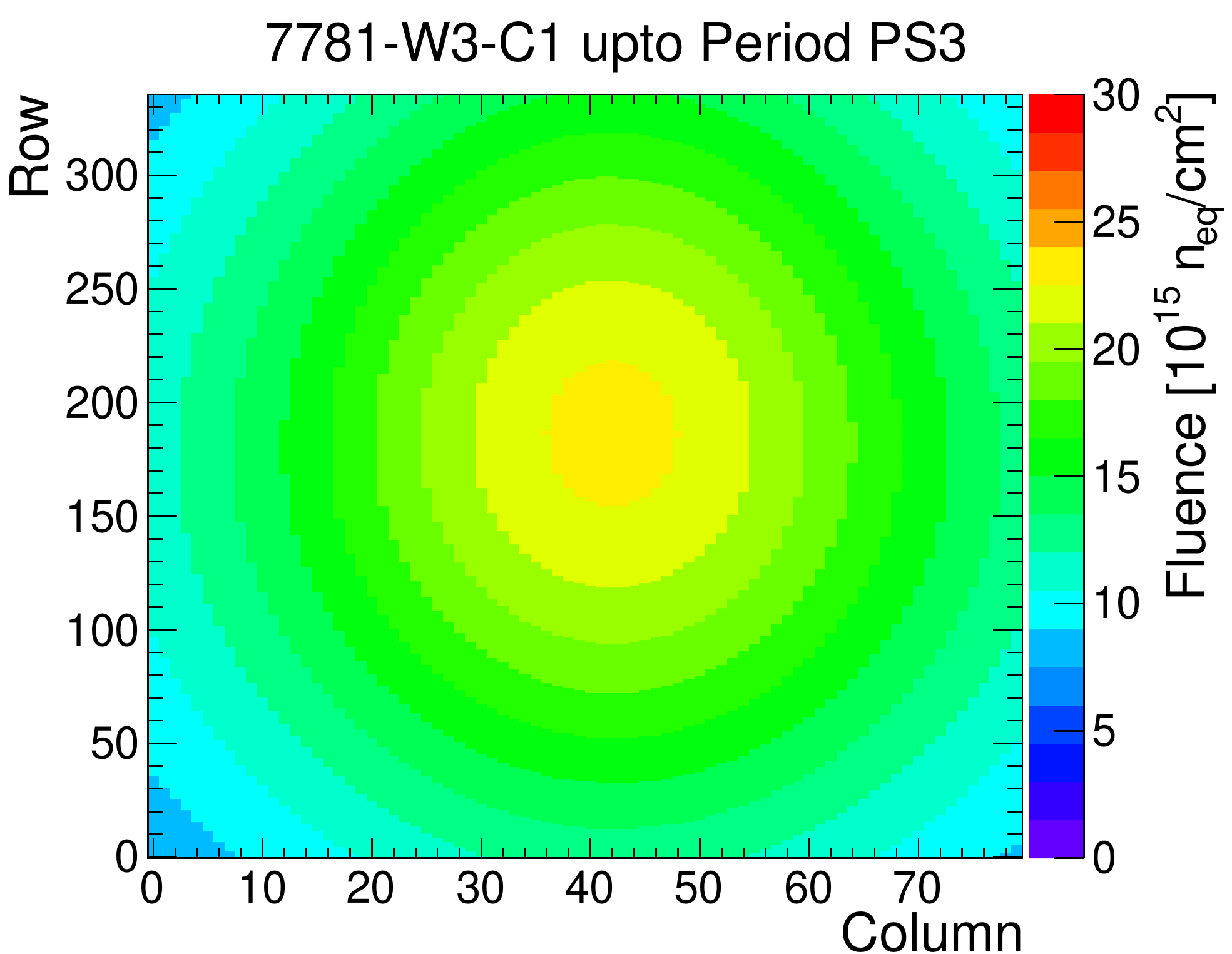}
	\includegraphics[width=3.7cm]{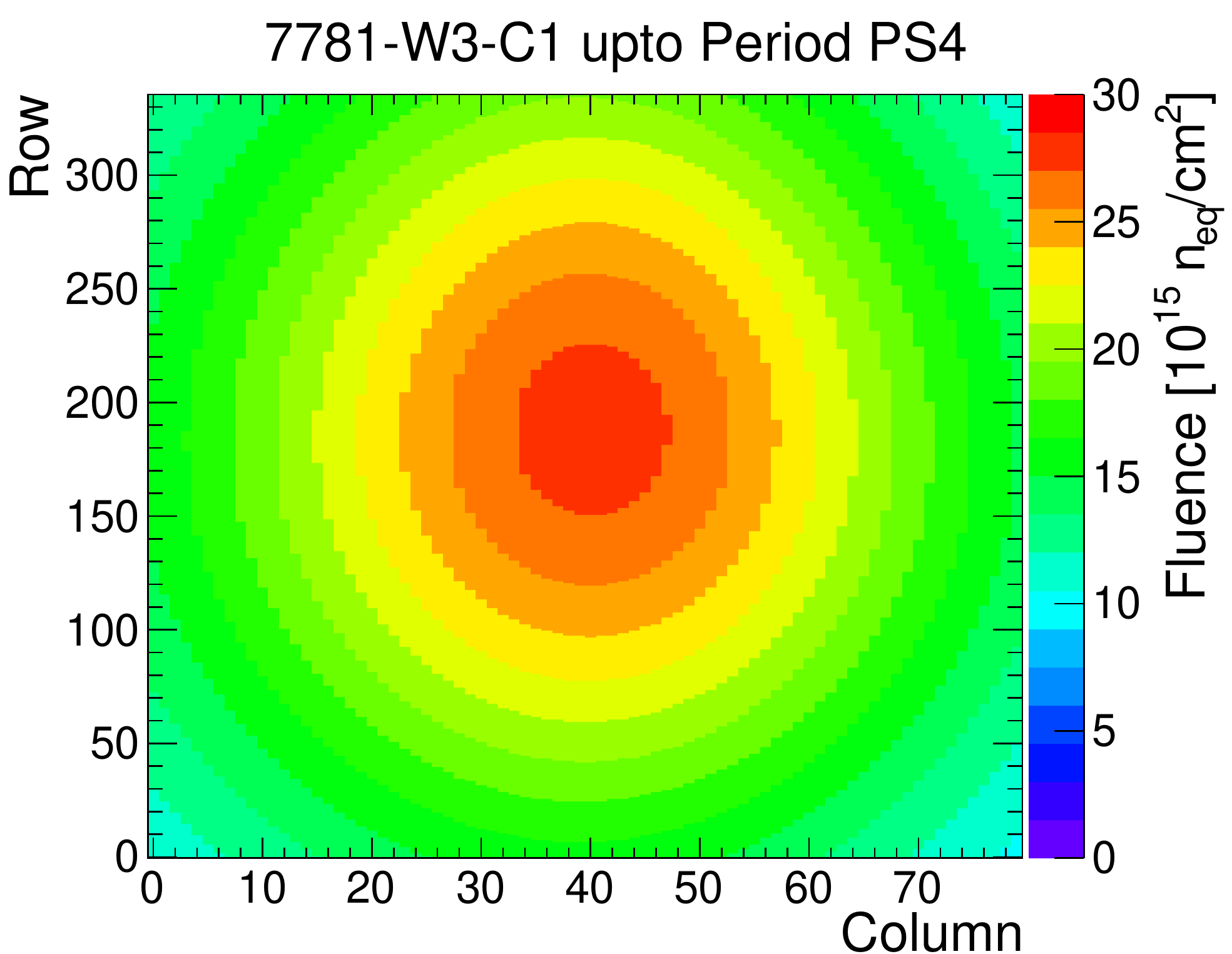}
	\includegraphics[width=3.7cm]{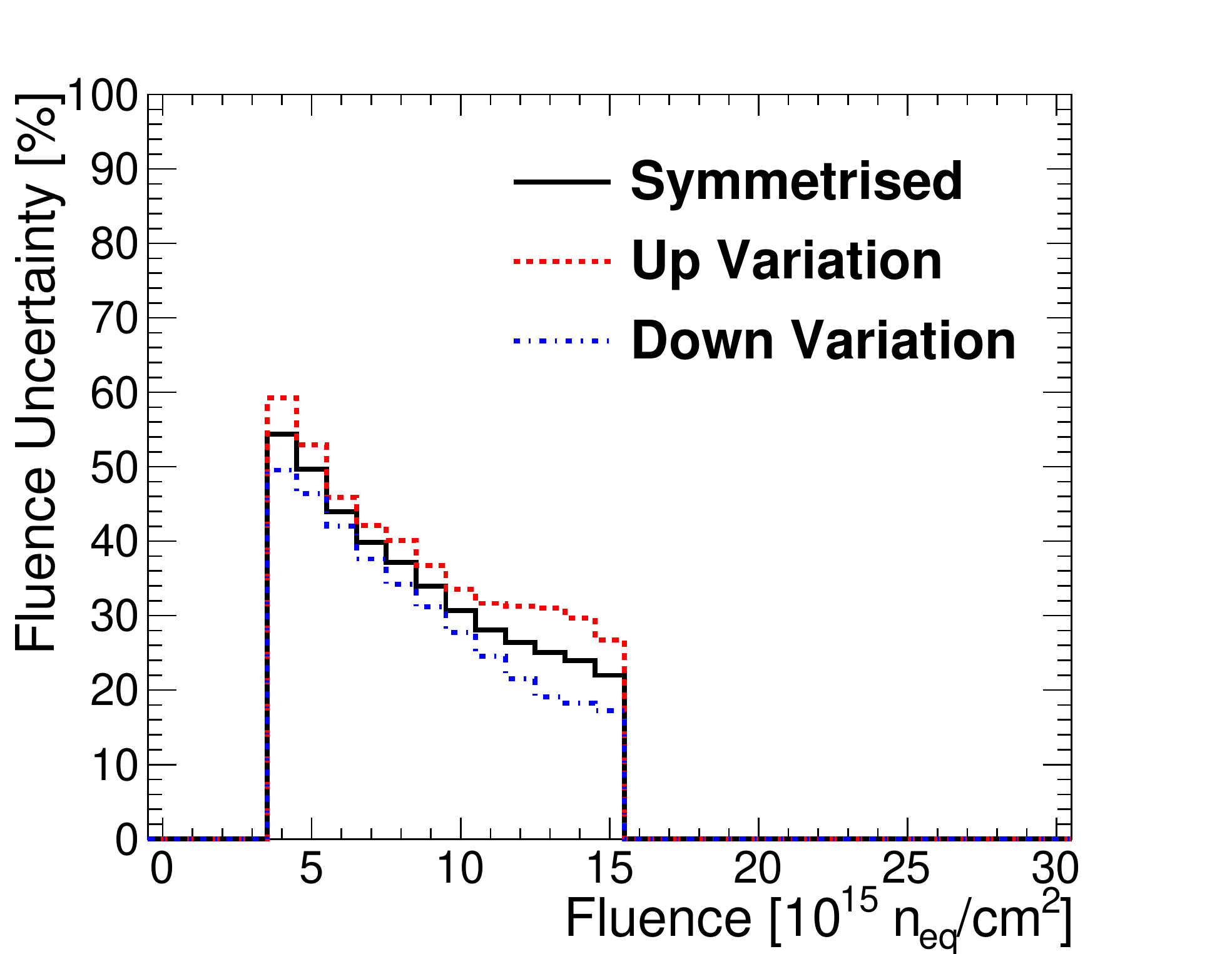}
	\includegraphics[width=3.7cm]{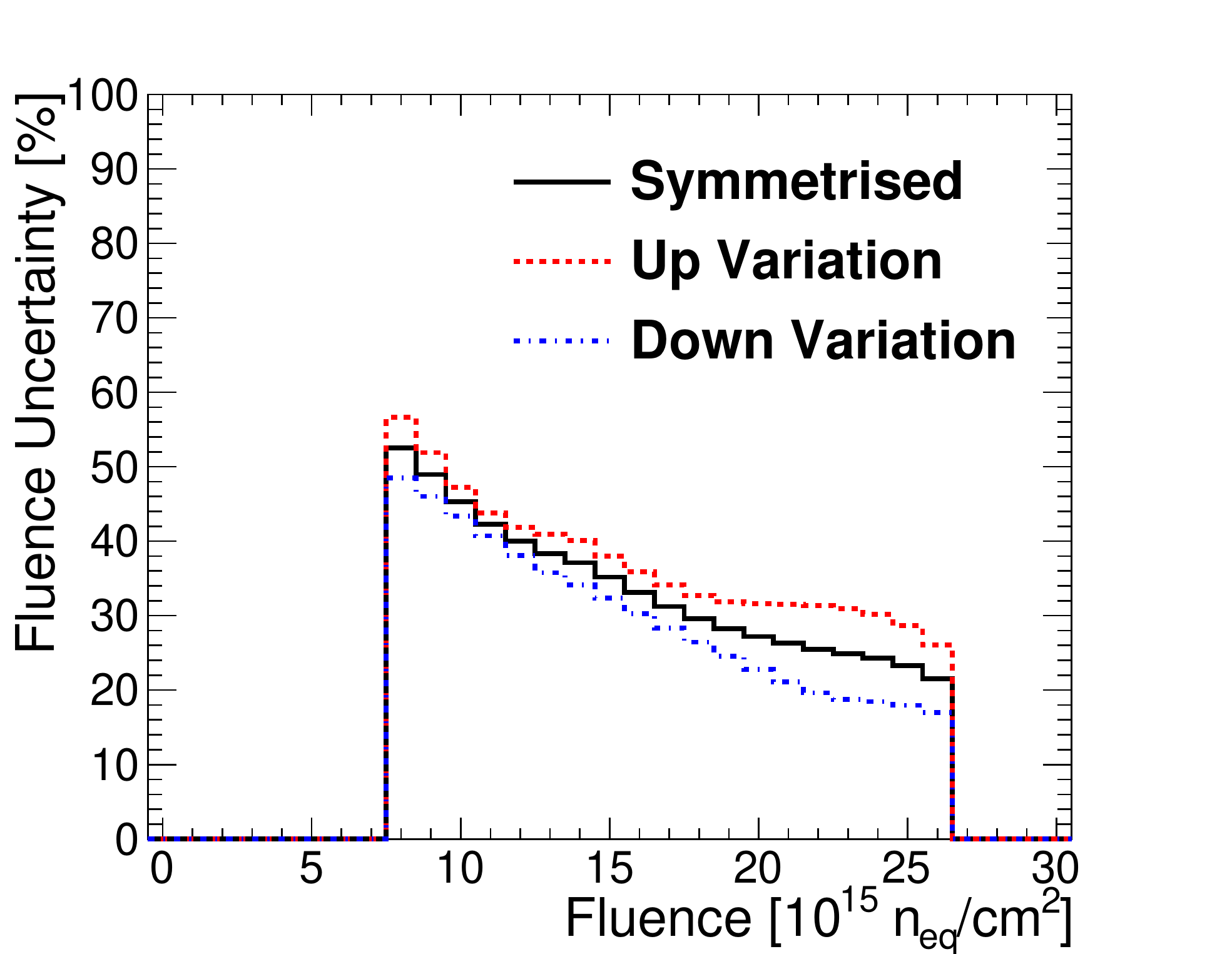}
	\includegraphics[width=3.7cm]{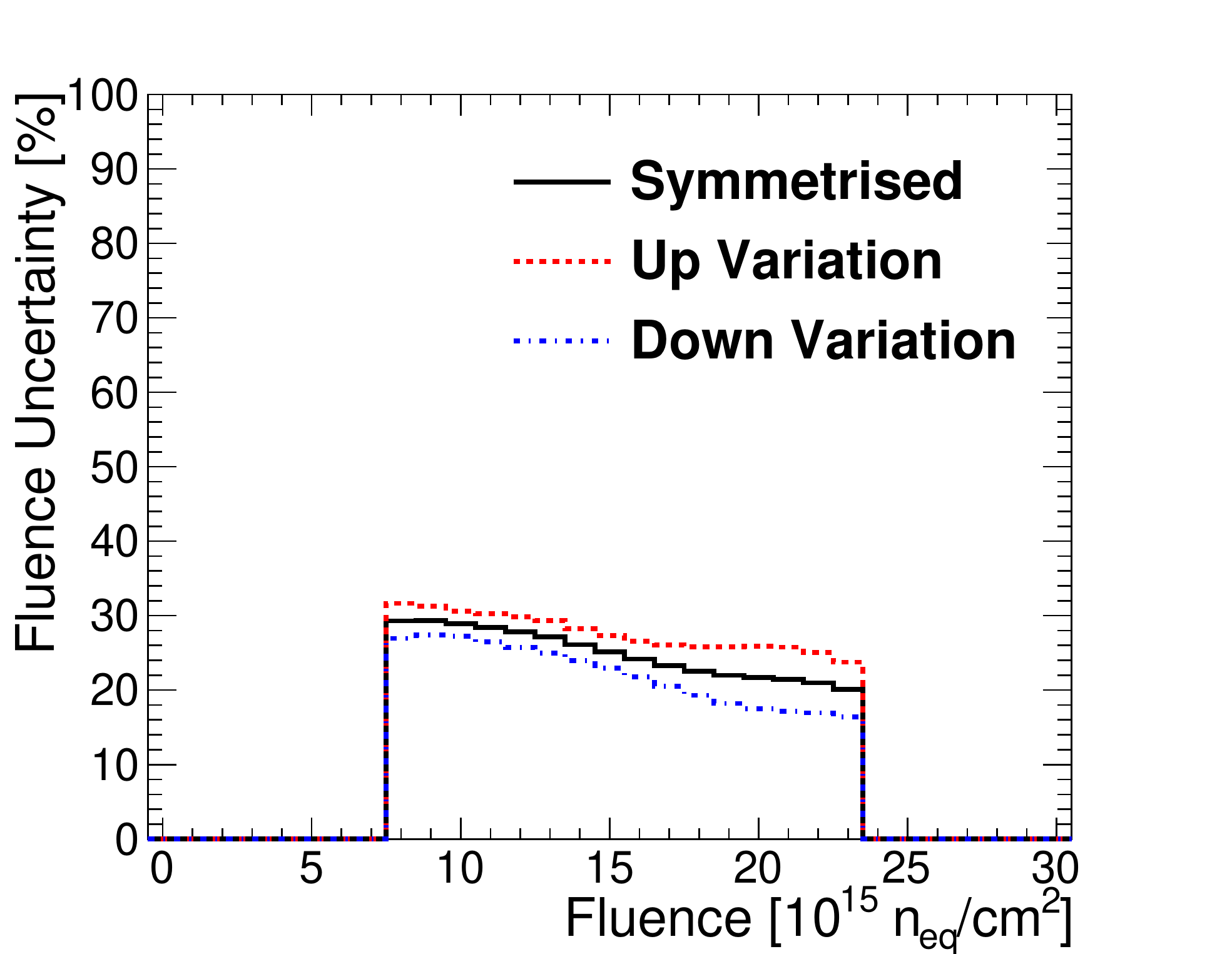}
	\includegraphics[width=3.7cm]{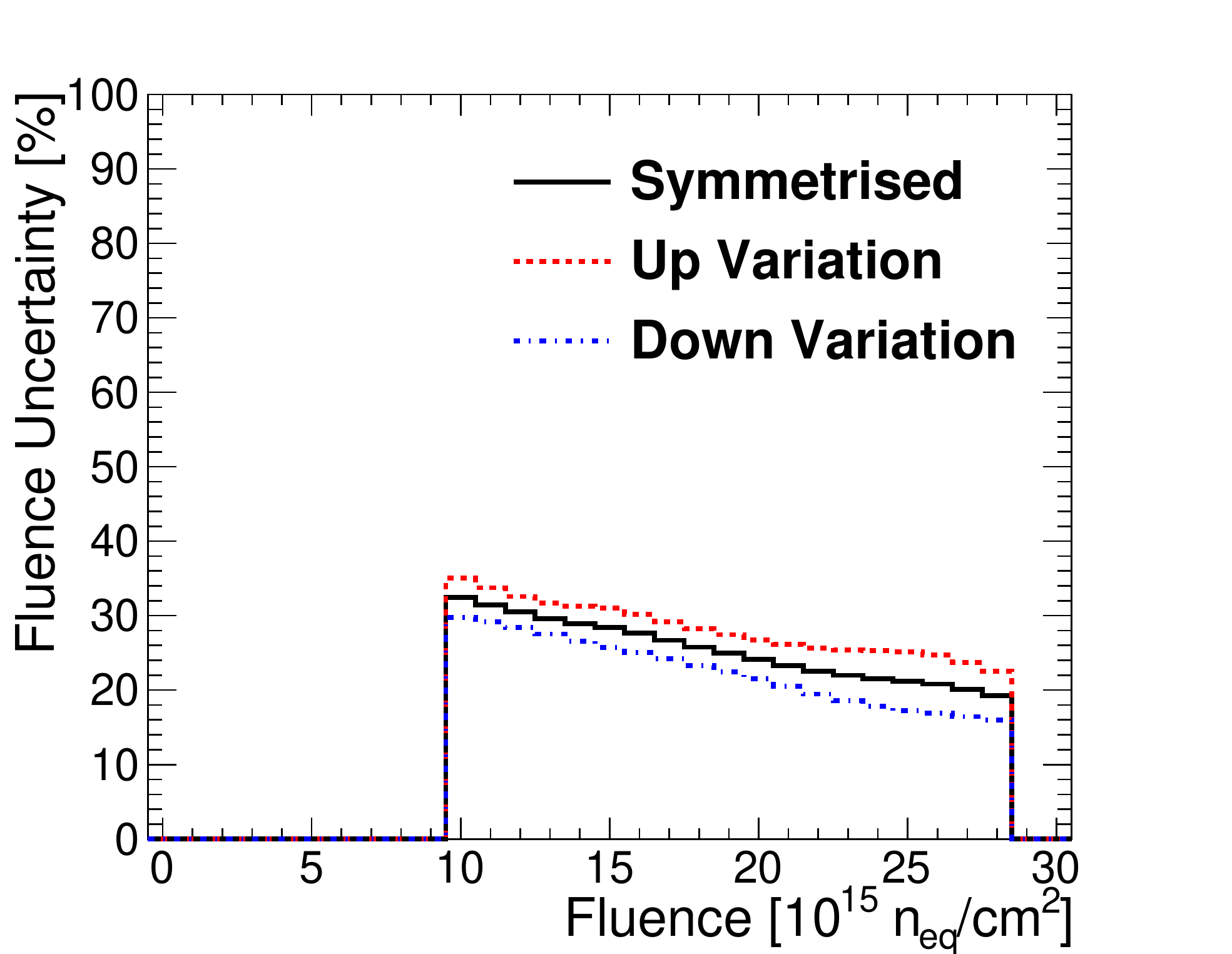}
	\caption{Top: integrated fluence maps as a function of pixel column and row numbers for all combinations of devices and PS irradiation periods studied. Bottom: corresponding systematic uncertainty as a function of fluence (up and down variations, as well as their symmetrised mean values, are shown).}
	\label{fig:fluenceMaps}
\end{figure}

Thus, to reach higher fluences beyond the baseline scenario and to test also higher-energy protons, five periods of irradiations (PS1--5) have been performed at the CERN-PS IRRAD facility with 23\,GeV protons with a Gaussian beam profile between 12 and 20\,mm FWHM. Hence, the fluence is non-uniformly distributed over the detector area. This makes it on the one hand possible to study a wide range of fluences on one single pixel detector as already demonstrated in reference~\cite{bib:CNMIBLgenAndSmallPitch2}. However, on the other hand, it also necessitates the individual determination of the fluence profile and an estimate of its uncertainties. The overall average fluence normalisation has been obtained from gamma spectroscopy of an activated 20$\times$20\,mm$^{2}$ aluminium (Al) foil for each period and device. The profile has been determined firstly using beam profile monitors in the beam line for all periods. For the two periods PS3 and PS4, it has been also determined by cutting the Al foil in $3\times3$ equal sub-foils, which have been measured individually (see figure~\ref{fig:fluenceMapsAl}) and fitted with a Gaussian. The two methods were found to be consistent with each other. The sub-foil method also allows to determine the centre of the beam with respect to the foil. As can be seen from figure~\ref{fig:fluenceMapsAl}, the beam hit the foil centrally for PS3, but with an offset of 3.5\,mm and 2.1\,mm in x and y, respectively, for PS4. For PS1 and PS2 where no sub-foil method was performed, the beam centre has been determined in-situ on the pixel devices themselves using the fluence dependence of the noise, the threshold before tuning and the optimum chip parameter values after threshold tuning. This is only possible for the first irradiation step of each device since the effects of several irradiation steps overlay each other. The resulting integrated fluence distributions for each combination of studied pixel device and irradiation period are shown in figure~\ref{fig:fluenceMaps} top. The peak fluences reached range from 1.5$\times10^{16}$\,n$_{\mathrm{eq}}$/cm$^2$ to 2.8$\times10^{16}$\,n$_{\mathrm{eq}}$/cm$^2$. The maximum fluence quoted in the analysis below can be slightly lower due to large disconnected regions of device W4-C1 (roughly the upper 60\% of the sensor) and binning effects. Device W3-C1 has been irradiated further in period PS5 to a peak fluence of 3.1$\times10^{16}$\,n$_{\mathrm{eq}}$/cm$^2$, but has not been measured yet in a beam test. 

Fluence uncertainties have been estimated by varying the following parameters: the fluence normalisation by 7\% (PS) or 10\% (KIT) according to the uncertainties quoted by the irradiation centres; the beam centre by 1\,mm; the Al foil centre with respect to the device by 1\,mm; and the beam $\sigma$ by 1\,mm. Since these irregular systematic shape effects do not propagate following the standard uncertainty formula of Gaussian distributions, a very conservative estimate has been obtained using the most extreme deviation of all variation combinations. The resulting maximum deviation as a function of fluence is shown in figure~\ref{fig:fluenceMaps} bottom after each integrated fluence period. For a pure PS irradiation (W4-C1, PS1 and PS3), the uncertainty increases steeply from the highest fluence range with typically about 20\% to the lowest fluence range with up to 50\%. For W3-C1, this dependence is less strong due to the fact that the first irradiation step has been performed at KIT with a flat 10\% normalisation uncertainty.

The irradiation at PS has been performed at room temperature. Moreover, after each irradiation step, about 1--3 weeks of storage (and hence annealing) at room temperature were needed for some initial decay of short-lived isotopes before handling, which is included in table~\ref{tab:samplesIrradiations}.

The results of the small-pitch 3D pixel detectors are compared to previously irradiated and studied devices from the IBL generation of 50$\times$250\,$\mu$m$^{2}$ 2E pixel geometry~\cite{bib:IBLprototypes,bib:CNMIBLgenAndSmallPitch2}, which are included for completeness in table~\ref{tab:samplesIrradiations}.

\section{Experimental techniques: laboratory measurements and beam tests}
\label{sec:Measurements}

The irradiated small-pitch 3D FE-I4 pixel prototypes were measured in the laboratory and beam tests at the Super Proton Synchrotron (SPS) H6 beam line at CERN with 120\,GeV pions. The Reconfigurable Cluster Element High Speed Input Output 2 (RCE HSIO2) readout system~\cite{bib:RCE1} was used for communication and data transfer from and to the FE-I4 readout chip. The devices were cooled either in a temperature controlled climate chamber in the laboratory, in a freezer or in a custom-made cooling box in the beam test to temperatures estimated between -10 and -30$^{\circ}$C on-sensor (see section~\ref{sec:leakageCurrent}). 

For the reconstruction of reference tracks in the beam test, a EUDET-type telescope out of six MIMOSA planes and scintillators for triggering was used~\cite{bib:EUDET}, in combination with the reconstruction and analysis frameworks EUTelescope and TBmonII~\cite{bib:EUtel,bib:TBmon2,bib:ATLAStestbeams}. The devices under test (DUT) were placed in the middle between an upstream and downstream arm of the telescope out of 3 planes each. The reference tracks are extrapolated to the position of each DUT with an estimated precision of about 4\,$\mu$m. The sensitive telescope area is about 1.1x2.1\,cm$^2$, so that only part of the the FE-I4 DUT area could be covered. To take into account that only 20\% of the small-pitch 3D prototype sensor region was active, as explained in section~\ref{sec:Devices}, the tracks used in the analysis were constrained to the ones passing through that active area inside each FE-I4 pixel. In order to also avoid smearing effects due to the telescope resolution, the region of interest (ROI), for which average results will be quoted in the following, was further reduced to the central 50$\times$50\,$\mu$m$^{2}$ of the two horizontally adjacent sensor pixels that were read-out (see figure~\ref{fig:3Dgeometry} bottom). Test beam measurements have been mostly done at perpendicular beam incidence with respect to the sensor surface (i.e. 0$^{\circ}$ tilt), as well as partially at a tilt of 15$^{\circ}$. In the analysis, hits in adjacent pixels of the DUTs are combined together to form a cluster. The cluster centre was obtained as the centre of gravity of the individual hit positions using the time-over-threshold (ToT) as provided by the FE-I4 readout as weights.

\section{General device performance and tuning}
\label{sec:generalPerformance}

The first tests were related to the communication with the devices and their ability to be tuned and calibrated. This is especially important since the FE-I4 readout chip is only qualified up to the IBL radiation doses and fluence levels of about 2.5\,MGy and 5$\times10^{15}$\,n$_{\mathrm{eq}}$/cm$^2$~\cite{bib:IBLprototypes}. Hence, in these irradiations the chip was pushed significantly beyond its specifications. Nevertheless, as can be seen in table~\ref{tab:samplesIrradiations}, in many cases the device was responsive and tunable after irradiation and could be measured in beam tests. However, in four cases communication was not possible, two times in the first fluence steps and two times after previously successful irradiations and testing. This could be either due to chip radiation damage or related to failures of the device board that was also partly irradiated or due to handling damage before or after irradiation. The surviving modules were all of the 50$\times$50\,$\mu$m$^{2}$ 1E 3D cell type. 

They were tuned using the in-chip charge injection mechanism and global as well as per-pixel chip registers to low thresholds of nominally 1000 or 1500\,e$^{-}$ and a time-over-threshold (ToT) of 10 clock cycles (in units of 25\,ns) for an injected charge of 20\,ke$^{-}$ (referred to as 10ToT@20\,ke$^{-}$). For some devices at high fluences (W4-C1, PS3 and W3-C1, PS4), only a ToT tuning to 10ToT@10\,ke$^{-}$ was possible. Especially the tuning of the non-uniformly irradiated devices at high fluences was a challenge since the non-uniform chip radiation damage required also the chip register values to be set as a function of pixel position. Nevertheless, in most cases satisfactory results were obtained after a few iterations using the standard RCE tuning algorithms, but fine-tuning manually certain chip register values. However, it should be noted that the exact performance (threshold and ToT dispersion, noise, number of noisy pixels) was fluctuating even for the same device for different tuning iterations and also depended on external conditions. The fraction of masked pixels that were discarded in the beam test readout or offline analysis due to either non-responsiveness or a too high dark count rate (so-called "hot" or "noisy" pixels), was typically a few \% in the centre of the devices. Sometimes the edge pixels at high voltages turned out to be noisier, in which case they were masked. With the exception of a strong increase of noisy pixels at very high voltages (higher voltages than needed for 97\% hit efficiency), 
no systematic trend on e.g. fluence and voltage in intermediate ranges was observed. 

Since this first study at such high fluences is mainly dedicated to the development of radiation-hard pixel sensor technology, whereas such effects are highly depending on the readout chip performance, its radiation damage and its exact tuning, a detailed further systematic study has not been performed on these small-pitch 3D FE-I4 prototypes. It will be carried out in the future, in combination with the new RD53A readout chip that was optimised to be operated at high fluences and low thresholds, once available. In the following, problematic pixels are simply masked and events with tracks through them or their neighbouring pixels are discarded from analysis.


\section{Leakage current and operation temperature}
\label{sec:leakageCurrent}

The sensor leakage current is an important parameter since it influences the noise and power dissipation and can damage the chip, if too high. At low fluences, it typically reaches a plateau value after full depletion, which increases linearly with irradiation fluence. However, at high fluences the behaviour is different and no plateau is observed, partly due to high electric fields leading to charge multiplication.

\begin{figure}[hbtp]
	\centering
	\includegraphics[width=7.5cm]{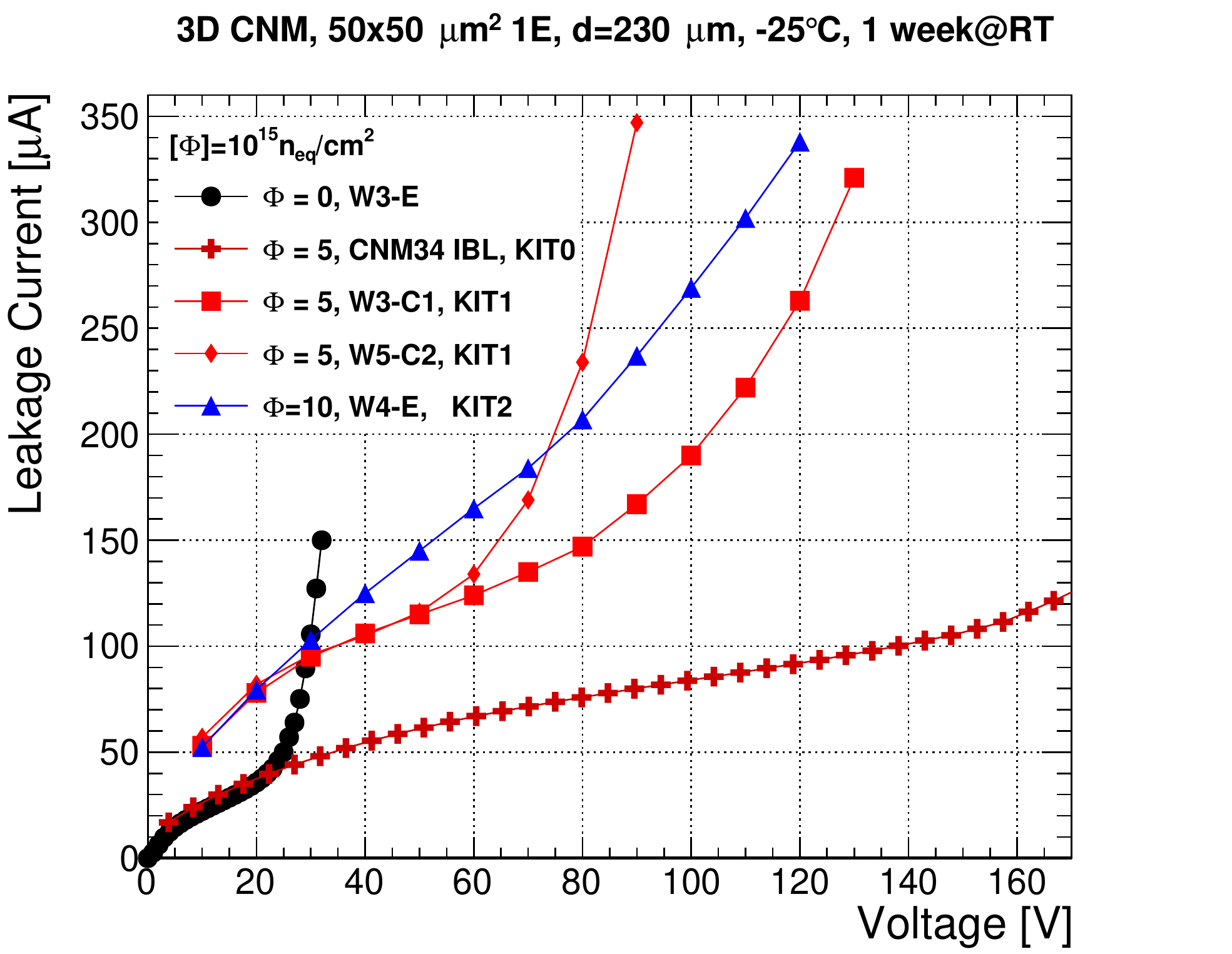}
	\includegraphics[width=7.5cm]{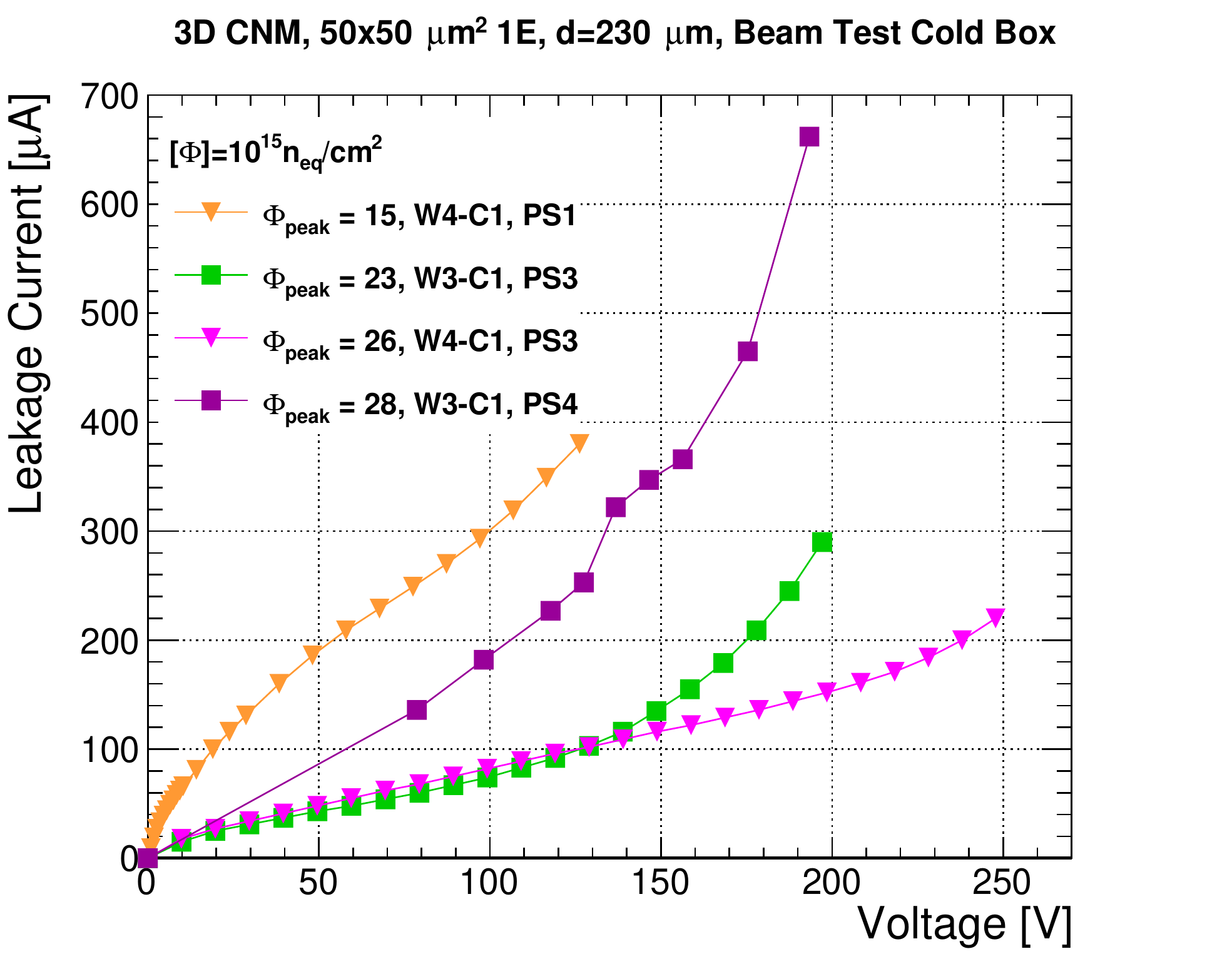}
	\caption{Leakage current as a function of voltage. Left: before and after uniform KIT irradiation, measured in a controlled climate chamber set to -25$^{\circ}$C with readout chip off. Right: after non-uniform PS irradiation in the beam test cold box.}
	\label{fig:Ileak}
\end{figure}

For the uniformly irradiated devices, a detailed study has been performed in a temperature controlled climate chamber with the readout chip off to avoid heating up the sensor. The default set temperature was the ITk baseline of -25$^{\circ}$C. However, due to poor thermal contact of the sensor with the environment, at high leakage currents a self-heating effect of the sensor appeared. To minimise the effect, the voltage ramp was performed fast. For the final ITk application, a better thermal conductivity to the cooling media is expected, which prevents this effect. The results are shown in figure~\ref{fig:Ileak} (left) for the small-pitch prototypes before and after irradiation and a standard 50$\times$250\,$\mu$m$^{2}$ 2E IBL FE-I4 for comparison. The current before irradiation is typically 20--40\,$\mu$A before the breakdown voltage of 15--40\,V. It can be seen how the current (before breakdown) increases with fluence, but so does the breakdown voltage, which makes the devices operable at higher voltages after irradiation. The currents of the two different devices at 5$\times10^{15}$\,n$_{\mathrm{eq}}$/cm$^2$ agree at low voltages, but the breakdown voltage is different. Comparing to the standard FE-I4 device at the same fluence, it can be seen that the current at a fixed voltage is higher for the new small-pitch generation. A similar effect has been observed before in 3D strip detectors~\cite{bib:CNMsmallPitch1}. An explanation could be higher electric fields causing charge multiplication due to the smaller inter-electrode distance and column diameter. However, at the operation point where the benchmark efficiency of 97\% is reached (40\,V for small-pitch, 120\,V for IBL, see section~\ref{sec:Efficiency}), the current values are similar. Some of the irradiated devices were also measured at different temperatures, and the expected scaling of radiation-induced leakage current with temperature was observed.

For non-uniformly irradiated devices, such a systematic study has not been performed since in any case the conclusions would be limited due to the non-linearity and strong voltage dependence of the leakage current at high fluences. However for reference, to appreciate the general leakage current behaviour, figure~\ref{fig:Ileak} right shows their leakage currents during the beam test. In general, during the beam test measurements, the temperature was not well controlled and also the FE-I4 chip heats up the sensor substantially when switched on, which also depends on the data rate, the thermal contact of the sensor and other external factors. The temperature of the cooling box was typically set to -40 or -50$^{\circ}$C during measurements. For dedicated tests on few devices, the currents were remeasured at different set temperatures with chip off. From comparison to the current during operation, the on-sensor operation temperature could be estimated as between -10 to -30$^{\circ}$C for these cases. Deviations from the expected fluence dependence can originate from temperature fluctuations (including self-heating at high leakage currents) or different annealing states.

\section{Time-over-threshold}
\label{sec:ToT}

The collected charge is another important sensor parameter for the device performance, since it directly impacts the signal height and hence the fraction of events that are above the detection threshold, i.e. hit efficiency. The FE-I4 chip measures the time-over-threshold (ToT), which is related to the charge, however with limited resolution (4 bits) and including non-linear effects that become stronger further away from the target charge value of the tuning point (here typically 10ToT@20\,ke$^{-}$). More precise measurements of the charge collected in small-pitch 3D sensors are carried out on neutron and proton irradiated 3D strip test structures from the same wafers with the ALIBAVA readout system~\cite{bib:neutronStripQ,bib:protonStripQ}.

\begin{figure}[hbtp]
	\centering
	\includegraphics[width=7.5cm]{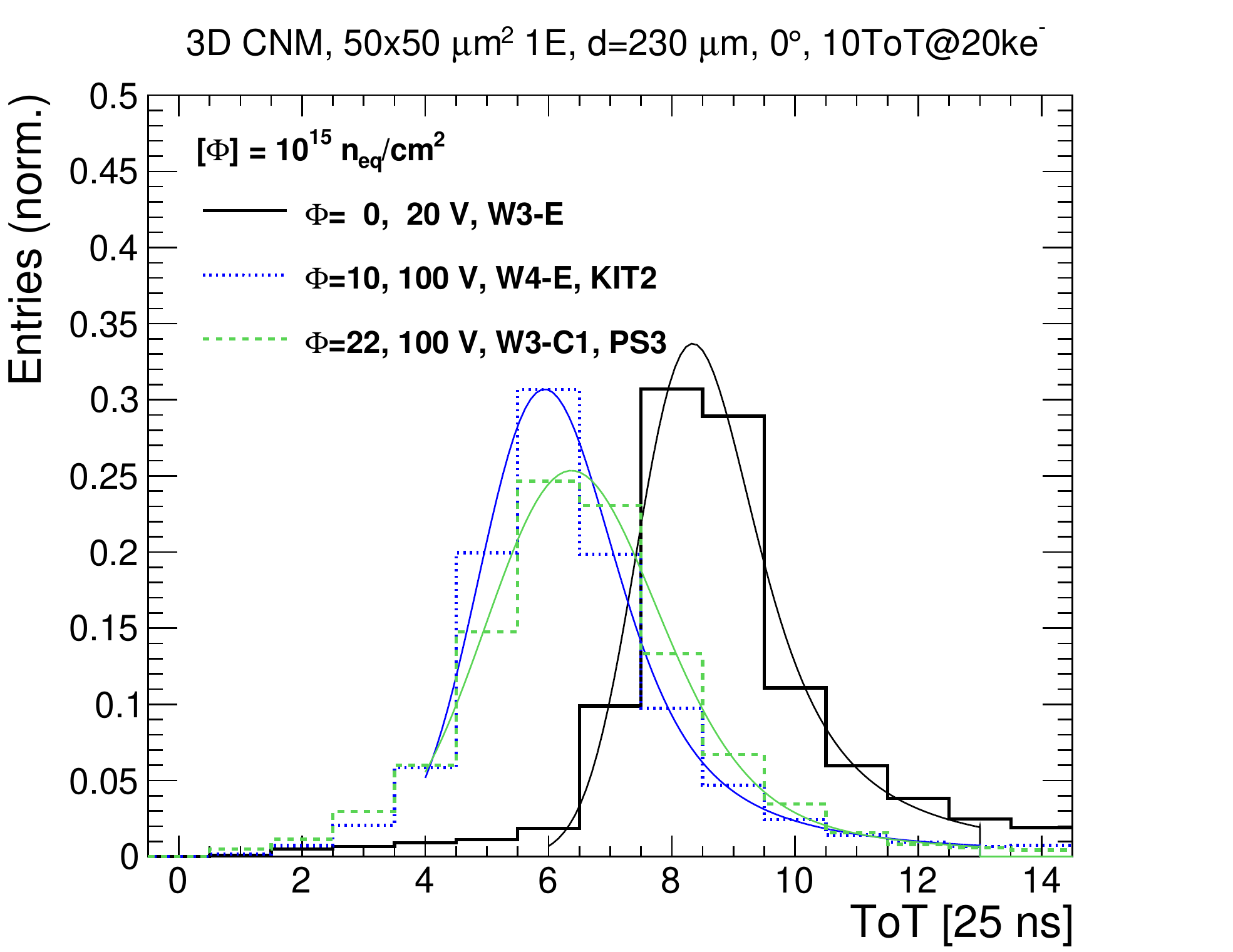}
	\includegraphics[width=7.5cm]{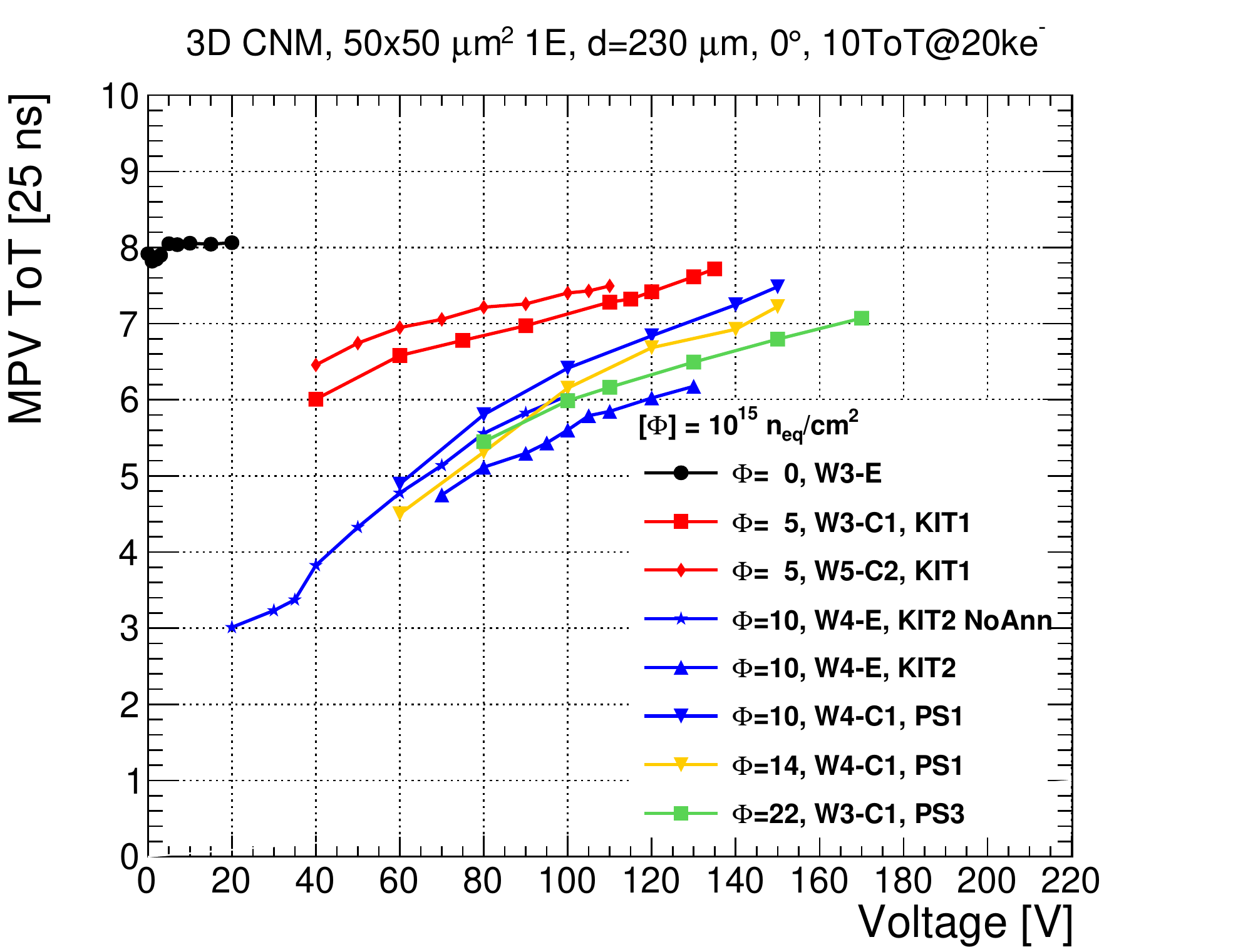}
	\caption{Left: cluster ToT distribution for different fluences. Right: cluster ToT MPV as a function of voltage. Statistical uncertainties are typically smaller than the marker size, systematic uncertainties (not shown for clarity) are estimated as 0.5\,ToT units. All measurements performed at perpendicular beam incidence.}
	\label{fig:ToT}
\end{figure}

\begin{figure}[hbtp]
	\centering
	\includegraphics[width=15cm]{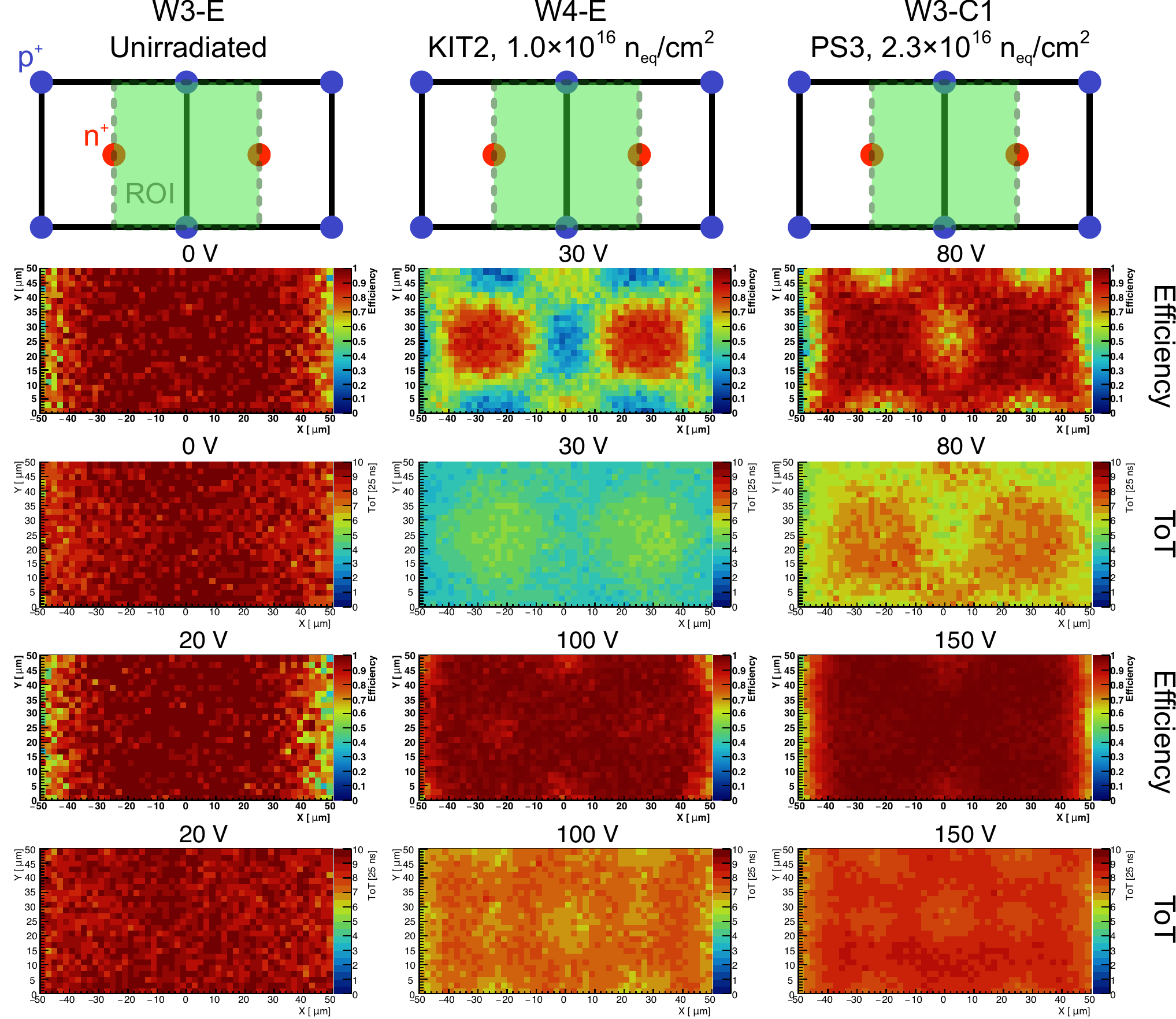}
	\caption{In-pixel average-ToT and efficiency maps restricted to the sensitive area of two adjacent 50$\times$50\,$\mu$m$^{2}$ sensor pixels connected to the readout chip (see top sketches) for selected samples, fluences and voltages.}
	\label{fig:inPixelMaps}
\end{figure}

Figure~\ref{fig:ToT} left shows the cluster ToT distributions in the central ROI before irradiation at 20\,V and after irradiation at 100\,V. The data are fitted with a Landau-Gauss distribution, and its most probable value (MPV) is shown in figure~\ref{fig:ToT} right as a function of voltage. Only the devices that could be measured at the standard ToT tuning point of 10ToT@20\,ke$^{-}$ are shown. Before irradiation, already at 0\,V external bias voltage almost the full charge (ToT$\approx$8) is collected due to the small 3D inter-electrode distance and hence almost immediate full depletion purely from the built-in voltage of the p-n junction ($\approx$0.7\,V). After irradiation, the ToT decreases at a fixed voltage due to trapping up to a fluence of 1.0$\times10^{16}$\,n$_{\mathrm{eq}}$/cm$^2$. However, the charge can be largely recovered by increasing the voltage: a ToT value of 6 is reached for all fluences studied at voltages between 40 and 120\,V. Again, this is an advantage of the small 3D inter-electrode distance and hence reduced trapping with respect to devices with larger inter-electrode distances. Possibly also charge multiplication effects as observed before in highly irradiated 3D strip sensors~\cite{bib:3Dmultiplication} play a role at higher voltages. The three measurements at 1.0$\times10^{16}$\,n$_{\mathrm{eq}}$/cm$^2$ agree within estimated systematic uncertainties of 0.5 ToT units. It is interesting to observe that the ToT seems not to decrease significantly further after 1.0$\times10^{16}$\,n$_{\mathrm{eq}}$/cm$^2$, but the precision is limited and the hit efficiency does show the expected decrease with fluence at a fixed voltage, as will be shown in section~\ref{sec:Efficiency}.
Figure~\ref{fig:inPixelMaps} shows the average-ToT in-pixel maps, which will be discussed together with the analogous efficiency in-pixel maps in the next section.


\section{Hit efficiency}
\label{sec:Efficiency}

\begin{figure}[hbtp]
	\centering
	\includegraphics[width=7.5cm]{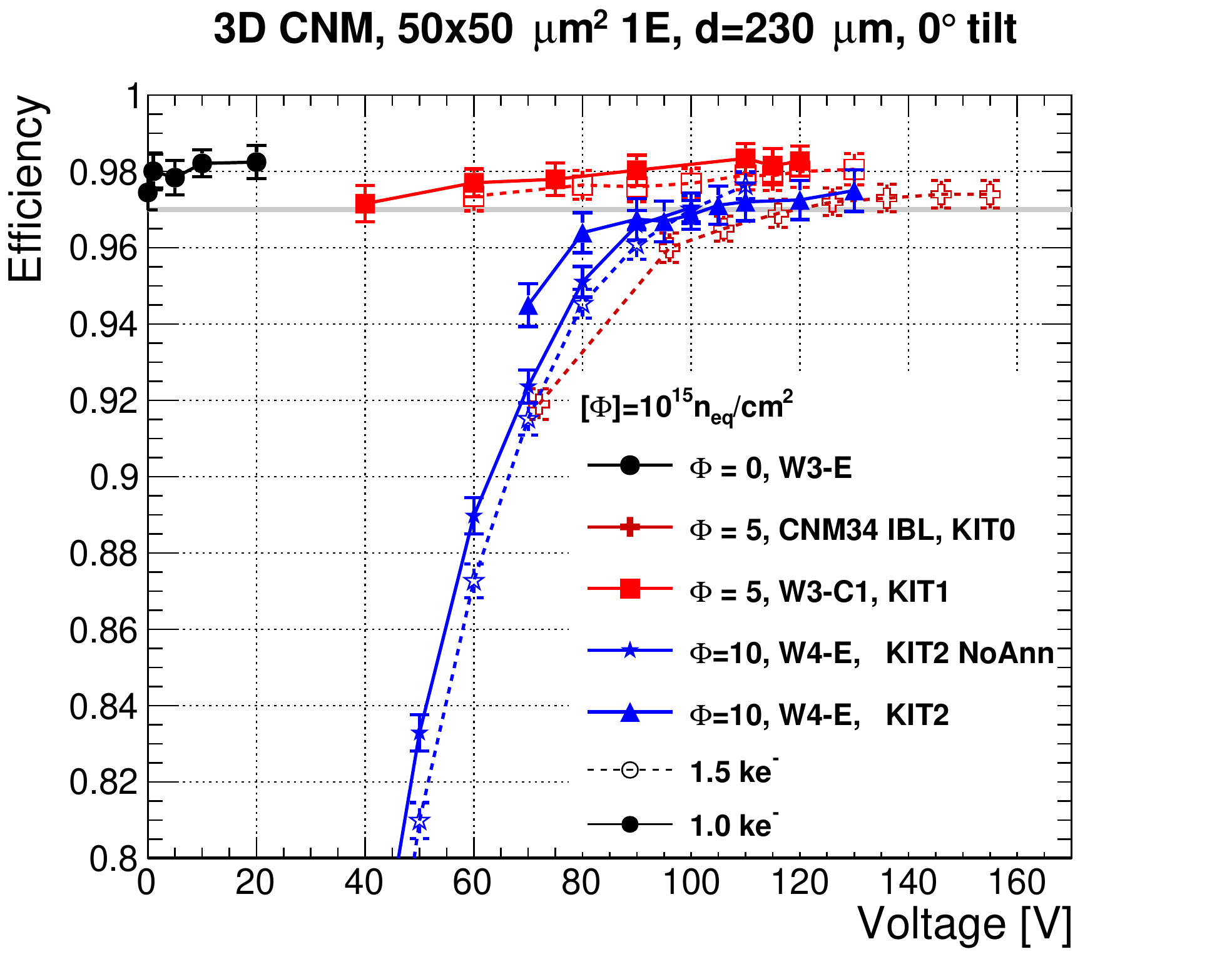}
	\includegraphics[width=7.5cm]{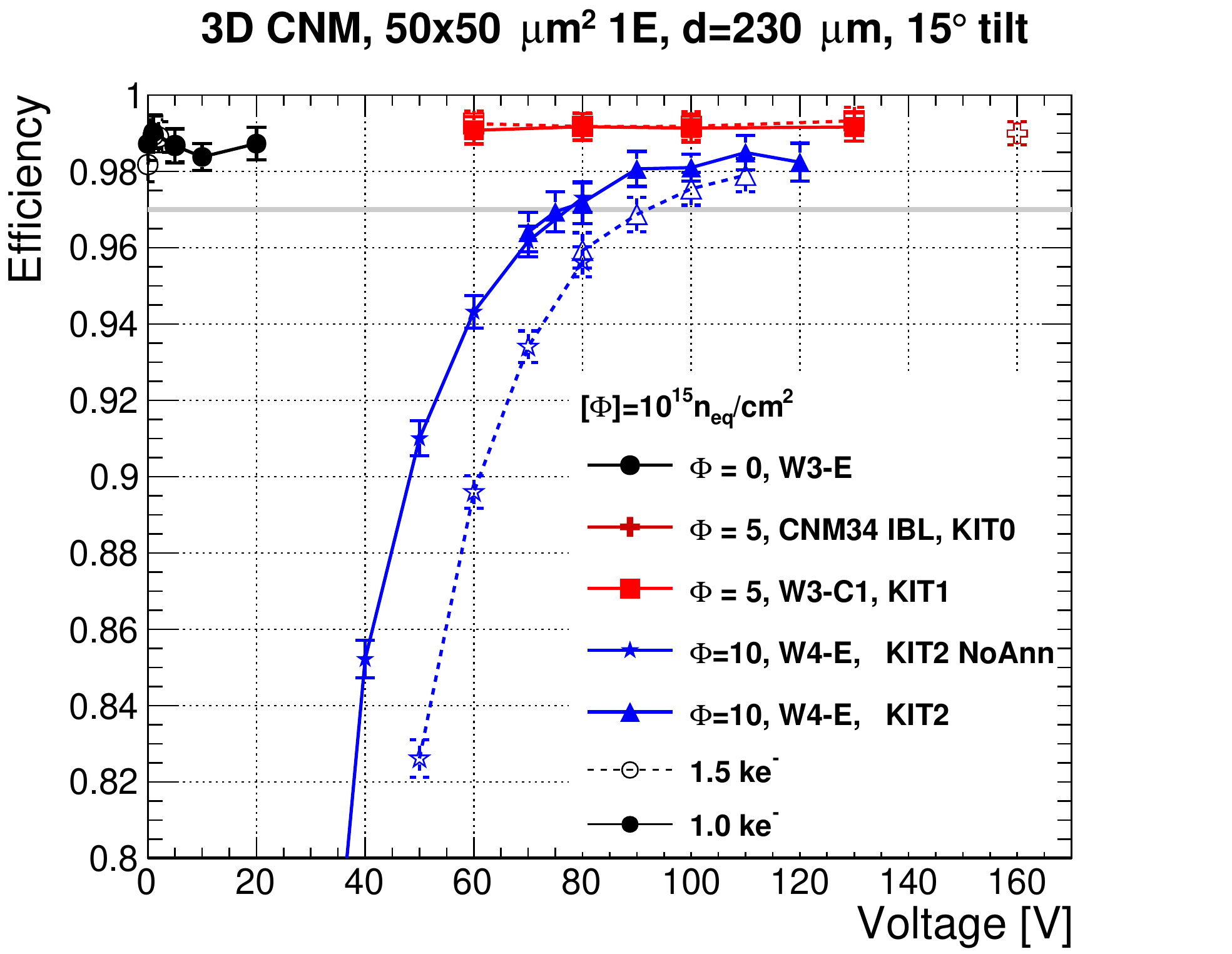}
	\caption{Hit efficiency as a function of voltage before and after uniform irradiation at KIT at 0$^{\circ}$ (left) and 15$^{\circ}$ (right) tilt. Open markers refer to a threshold of 1.5\,ke$^{-}$, full ones to 1.0\,ke$^{-}$. Uncertainties shown are statistical and systematic combined.}
	\label{fig:effUniform}
\end{figure}

The hit efficiency is defined as the fraction of events in which a particle passing through a DUT causes a recorded hit. Hence, it is highly influenced by the signal height (see section~\ref{sec:ToT}) and chip threshold setting. The minimum hit efficiency target for ITk is 97\% throughout the whole lifetime since lower values would cause problems for track pattern recognition~\cite{bib:ITkPixelTDR}. Experimentally, the efficiency is determined in beam tests by extrapolating the reference track position to the DUT and searching for a hit cluster in the surroundings within a matching distance of two times the FE-I4 pixel size, i.e. 100$\times$500\,$\mu$m$^{2}$. Systematic uncertainties as obtained from repeated measurements and analysis variations are estimated as 0.3\%, which is added in quadrature to the statistical ones.

Figure~\ref{fig:inPixelMaps} shows the efficiency in-pixel maps at 0$^{\circ}$ tilt, along with their average-ToT maps, of two adjacent sensor pixels at a relatively low and high voltage for three devices: before irradiation at 0 and 20\,V, after uniform irradiation to 1.0$\times10^{16}$\,n$_{\mathrm{eq}}$/cm$^2$ at 30 and 100\,V, and after a non-uniform peak fluence\footnote{The map is the average over all pixels in the measured sensor area, hence the resulting average fluence is less than the peak.} of 2.3$\times10^{16}$\,n$_{\mathrm{eq}}$/cm$^2$ at 80 and 150\,V. Before irradiation, the ToT and efficiency is observed to be very uniform and high, even at 0\,V. Only at the edges at $x=\pm$50\,$\mu$m close to the insensitive area, the efficiency is reduced due to an artifact, namely the telescope resolution smearing. After irradiation, at low voltages the ToT is higher around the 3D columns, especially the n$^{+}$ junction columns, leading to relatively high efficiencies there, whereas the area between two p$^{+}$ columns shows lower ToT and efficiencies due to larger drift distances and smaller electric fields. However, as can be seen, at high voltages the absolute value as well as uniformity of both parameters are restored. Only small effects are seen from the partly insensitive 3D columns, which sometimes cause locally low charge and inefficiencies at 0$^{\circ}$ tilt~\cite{bib:3D,bib:IBLprototypes}. This effect is less pronounced here than for previous generations due to the fact that these columns are non-fully passing through, the diameter has been reduced with respect to the IBL generation by 2\,$\mu$m down to nominally 8\,$\mu$m, and the columns are narrower at the tip (see section~\ref{sec:Devices}). Moreover, the telescope resolution of about 4\,$\mu$m is close to the column diameter and is hence diluting the effect.

Figure~\ref{fig:effUniform} shows the hit efficiency in the central ROI as a function of bias voltage at 0$^{\circ}$ (left) and 15$^{\circ}$ (right) tilt for an unirradiated device and the ones irradiated uniformly at KIT at different thresholds. The efficiencies are found to be slightly lower at 0$^{\circ}$ compared to 15$^{\circ}$ (by few permil to 1\%) due to the effect discussed above that at 0$^{\circ}$ tilt, a particle can pass exactly through the partly insensitive 3D column, which is not possible after tilting. Hence, 0$^{\circ}$ constitutes the worst-case scenario. The efficiencies at 1.5\,ke$^{-}$ are only slightly lower than at 1.0\,ke$^{-}$.
Before irradiation, the hit efficiency is already at its plateau value of 98-99\% even at 0\,V. This can be explained by the high level of depletion and signal just due to the built-in voltage of about 0.7\,V in combination with the small inter-electrode distance as already discussed in section~\ref{sec:ToT}. 
At 5$\times10^{15}$\,n$_{\mathrm{eq}}$/cm$^2$, the efficiency at 0$^{\circ}$ and 1.0\,ke$^{-}$ threshold reaches the ITk benchmark of 97\% already at 40\,V, increasing to 98\% at about 100\,V. At 15$^{\circ}$, a plateau efficiency of 99\% is observed. It can be seen that the efficiency of the new small-pitch generation with 50$\times$50\,$\mu$m$^{2}$ 1E pixels is significantly higher than the one for the IBL generation with 50$\times$250\,$\mu$m$^{2}$ 2E pixels, which reaches 97\% only at 120\,V, which is again explained by the smaller inter-electrode distance and hence less trapping. At 1.0$\times10^{16}$\,n$_{\mathrm{eq}}$/cm$^2$, the efficiency drops at a fixed voltage with respect to lower fluences, but 97\% is reached at about 100\,V (80\,V) for 0$^{\circ}$ (15$^{\circ}$) tilt at 1.0\,ke$^{-}$ threshold. The two measurements before and after annealing to 1week@RT agree well (in particular in the voltage needed for 97\% efficiency), except for low voltages (70--80\,V) at 0$^{\circ}$ tilt, for which the annealed device shows a few \% higher efficiency. It is not yet understood if this is a real annealing effect or an artifact. More systematic annealing studies need to be carried out in the future.

\begin{figure}[hbtp]
	\centering
	\includegraphics[width=7.5cm]{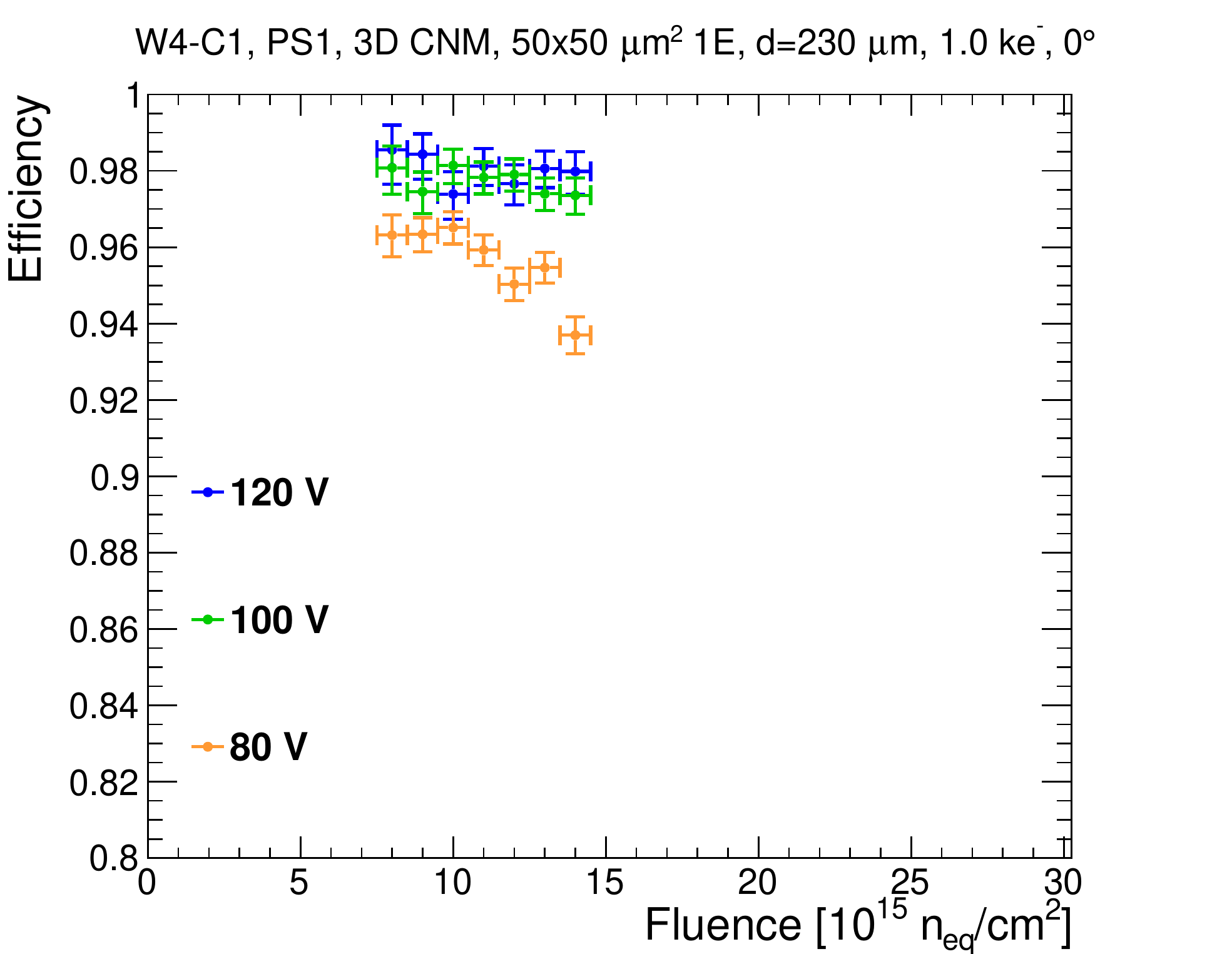}
	\includegraphics[width=7.5cm]{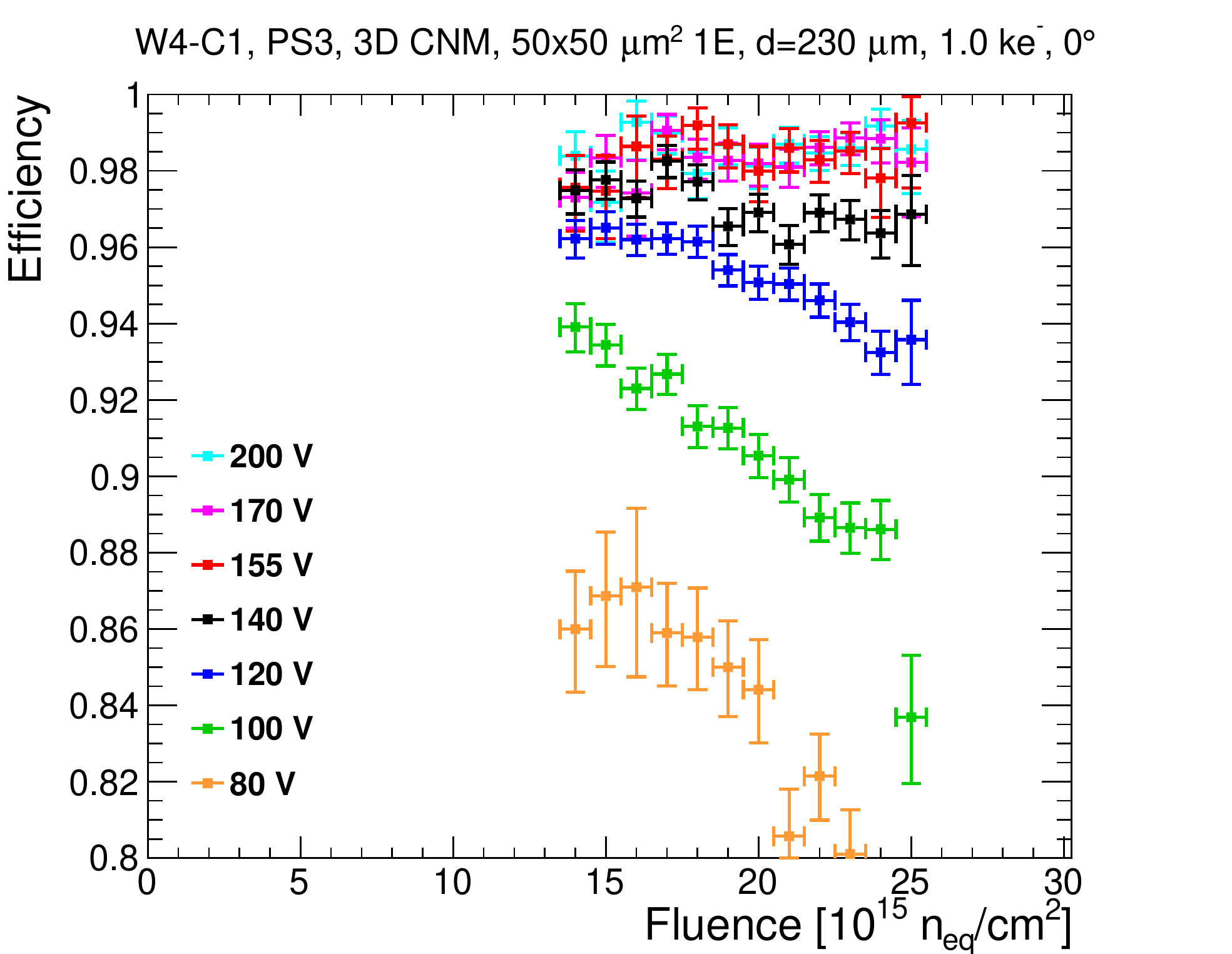}
	\includegraphics[width=7.5cm]{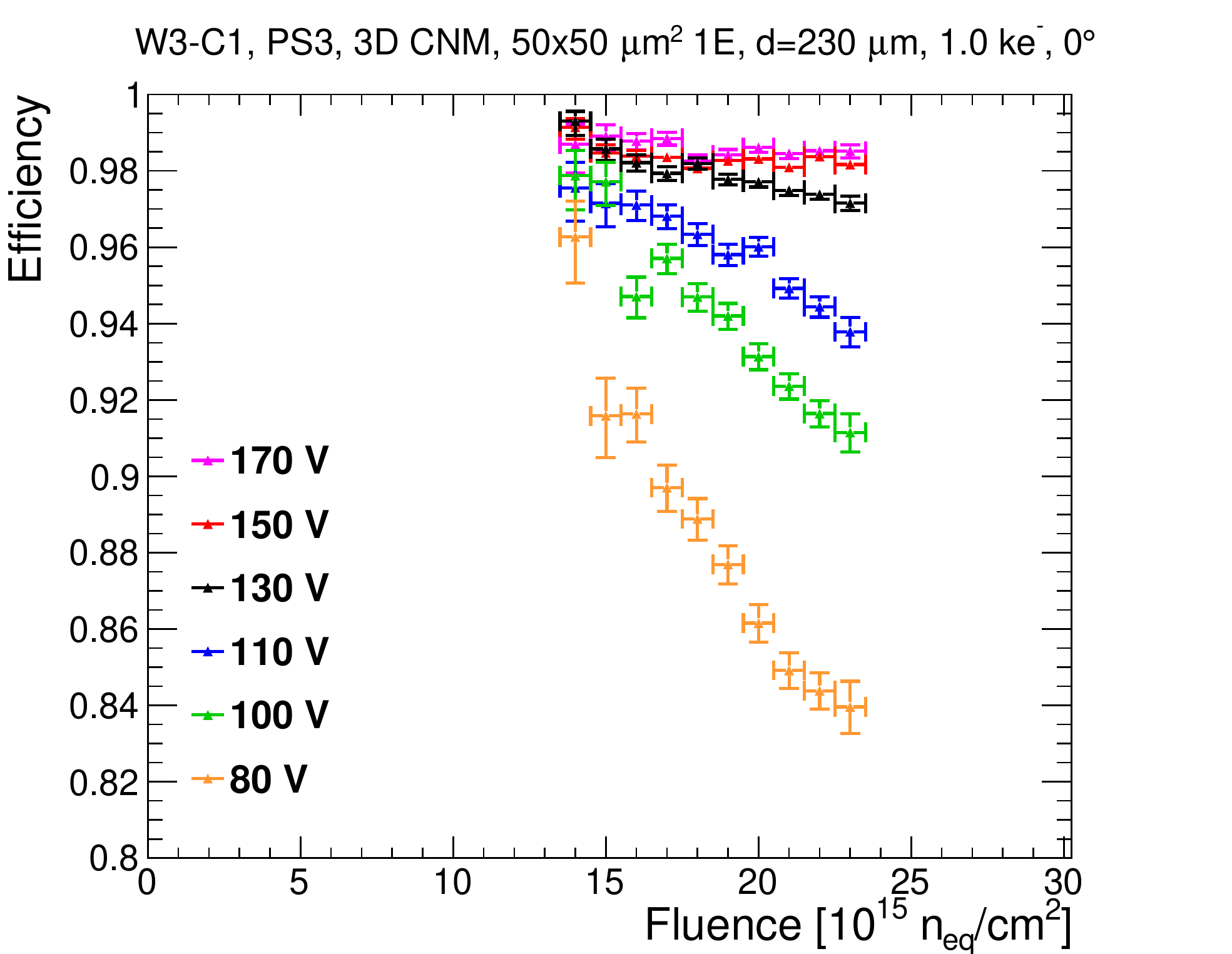}
	\includegraphics[width=7.5cm]{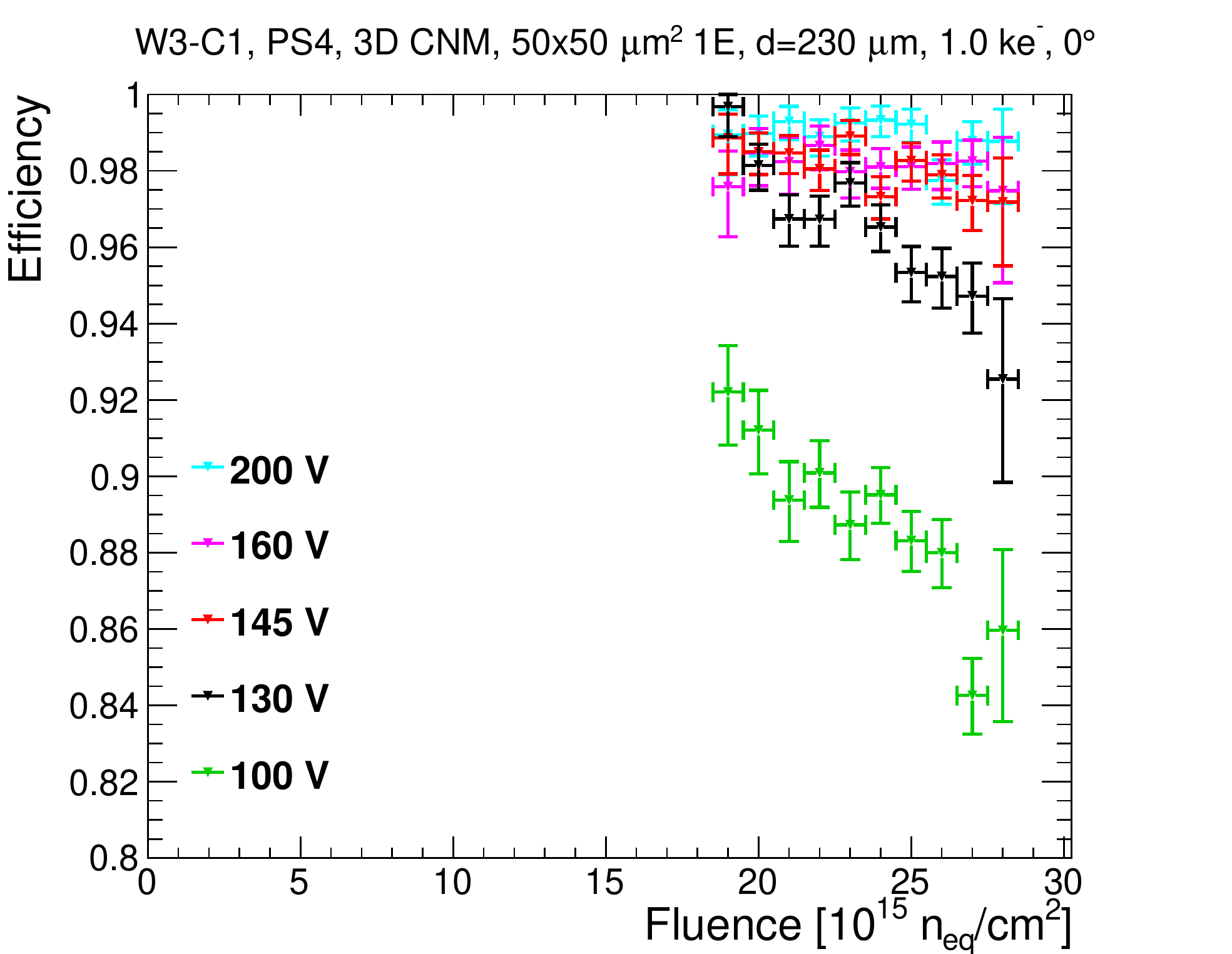}
	\caption{Hit efficiency as a function of fluence for different voltages at 0$^{\circ}$ and 1.0\,ke$^{-}$ obtained from one single pixel device: W4-C1 (top) and W3-C1 (bottom) at two different irradiation periods (left/right). The bars shown refer to combined statistical and systematic uncertainties for the efficiency and the bin size for the fluence. The systematic fluence uncertainties as displayed in figure~\ref{fig:fluenceMaps} are larger than each bin and are not shown for visibility.}
	\label{fig:effNonUniform}
\end{figure}

\begin{figure}[hbtp]
	\centering
	\includegraphics[width=7.5cm]{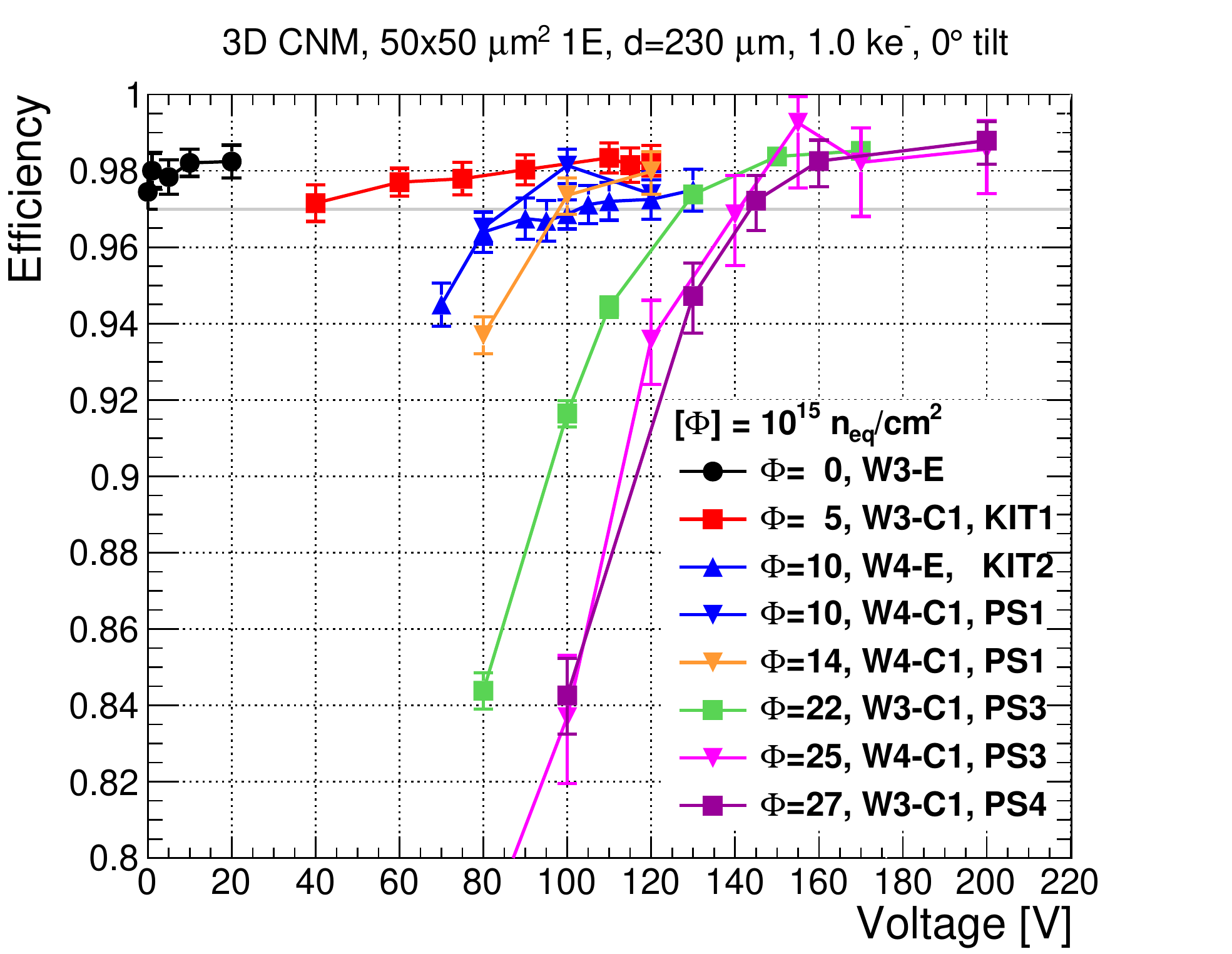}
	\includegraphics[width=7.5cm]{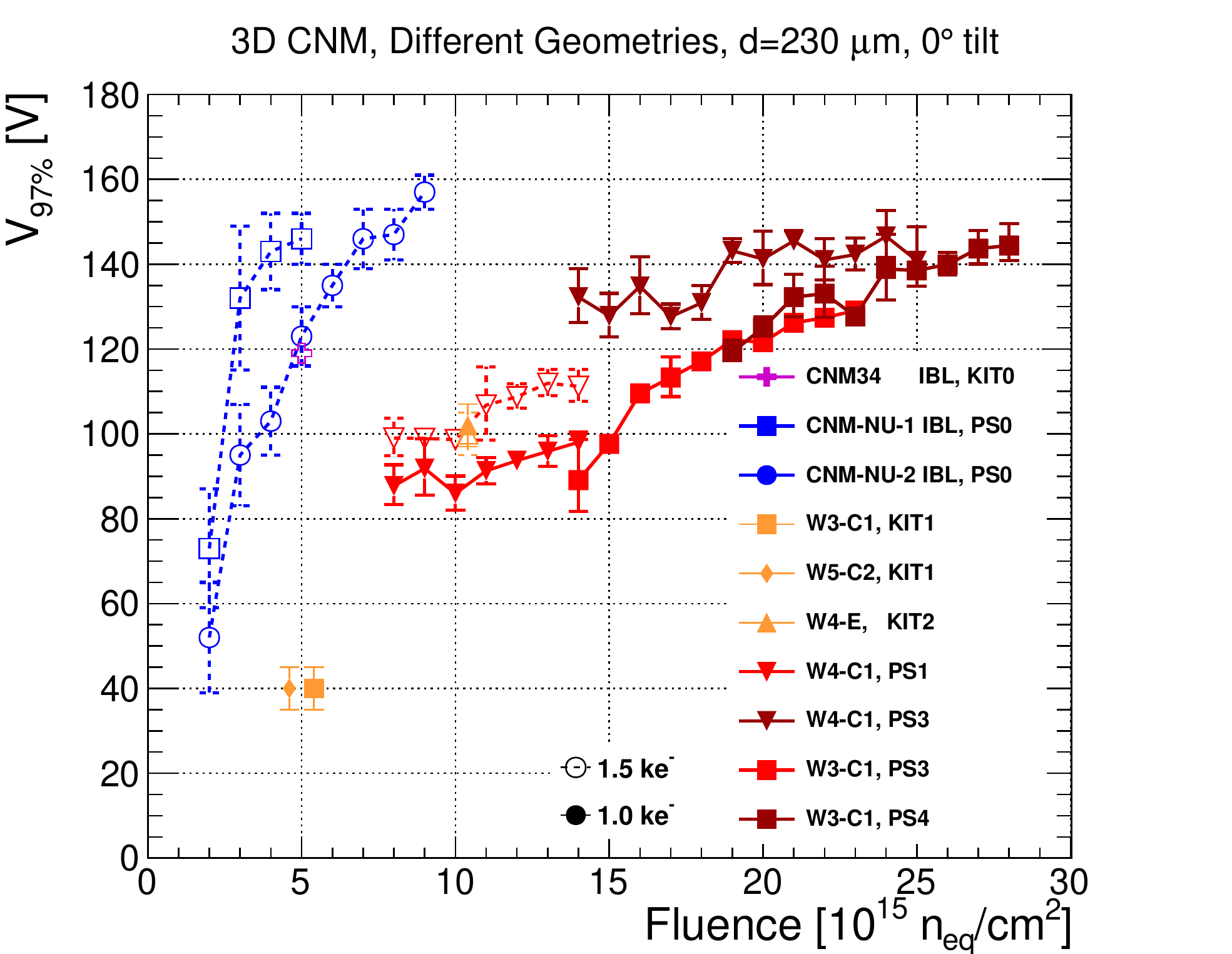}
	\caption{Left: hit efficiency as a function of voltage for different fluences at 0$^{\circ}$ and 1.0\,ke$^{-}$. Right: V$_{97\%}$ as a function of fluence at 0$^{\circ}$ tilt for different devices. The small-pitch 50$\times$50\,$\mu$m$^{2}$ 1E devices (red/orange) are compared to the IBL 50$\times$250\,$\mu$m$^{2}$ 2E generation (blue/magenta). Open markers refer to a threshold of 1.5\,ke$^{-}$, full ones to 1.0\,ke$^{-}$. The uncertainties shown are statistical and systematic combined. The fluence uncertainty from figure~\ref{fig:fluenceMaps} is not displayed for visibility.}
	\label{fig:effV97Compilation}
\end{figure}

For the non-uniformly irradiated devices, figure~\ref{fig:effNonUniform} shows the efficiency as a function of fluence (on one single device) for different voltages at 0$^{\circ}$ and 1.0\,ke$^{-}$. At low voltages, the expected decreasing trend with fluence is observed. This gives confidence that the peak fluence positions, as explained in section~\ref{sec:Irradiations}, are reasonably well determined. At high voltages, no fluence dependence is seen as the efficiency is uniformly high and has reached the plateau value everywhere. The efficiency values at the same voltage and fluence agree reasonably well for the different devices and periods, except for W4-C1, PS3 in the low-fluence region, which is observed to be systematically lower than the others. However, this can be understood when taking the systematic fluence uncertainties into account, which increase significantly with decreasing fluence as discussed in section~\ref{sec:Irradiations}. Hence, for a compilation of the voltage dependence of the efficiency as shown in figure~\ref{fig:effV97Compilation} left, the results are evaluated at (or close to) the highest fluence of each device, where systematic fluence uncertainties are lowest. Also the results before and after uniform irradiation are included. At 1.0$\times10^{16}$\,n$_{\mathrm{eq}}$/cm$^2$, the curves for KIT and PS irradiation agree reasonably well. The expected fluence degradation at fixed voltage is observed. However, even at the highest fluence studied, a plateau efficiency of 98\% is obtained. The voltage at which the ITk benchmark efficiency of 97\% is reached, V$_{97\%}$, is evaluated from linear interpolation and shown in figure~\ref{fig:effV97Compilation} right as a function of fluence. As expected, a rising trend is visible, but even at 2.8$\times10^{16}$\,n$_{\mathrm{eq}}$/cm$^2$, V$_{97\%}$ does not exceed 150\,V for the new small-pitch generation. At the ITk baseline fluence of 1.3$\times10^{16}$\,n$_{\mathrm{eq}}$/cm$^2$ (assuming one replacement), V$_{97\%}$ is only about 100\,V. The new small-pitch generation is also compared to measurements of the IBL 50$\times$250\,$\mu$m$^{2}$ 2E generation~\cite{bib:CNMIBLgenAndSmallPitch2}. The significant improvement of the new generation is visible, which was discussed for 5$\times10^{15}$\,n$_{\mathrm{eq}}$/cm$^2$ (KIT) already above. The most radiation-hard planar pixel technology up to date with 100\,$\mu$m thin sensors was evaluated up to 1.0$\times10^{16}$\,n$_{\mathrm{eq}}$/cm$^2$, and V$_{97\%}$ was found to be about 500\,V~\cite{bib:planarPower}. This is more than 5 times higher than needed for small-pitch 50$\times$50\,$\mu$m$^{2}$ 1E 3D sensors, demonstrating the superior radiation hardness of the 3D technology, especially for small inter-electrode distances.


\section{Edge efficiency}
\label{sec:SlimEdge}

\begin{figure}[hbtp]
	\centering
	\includegraphics[width=7.5cm]{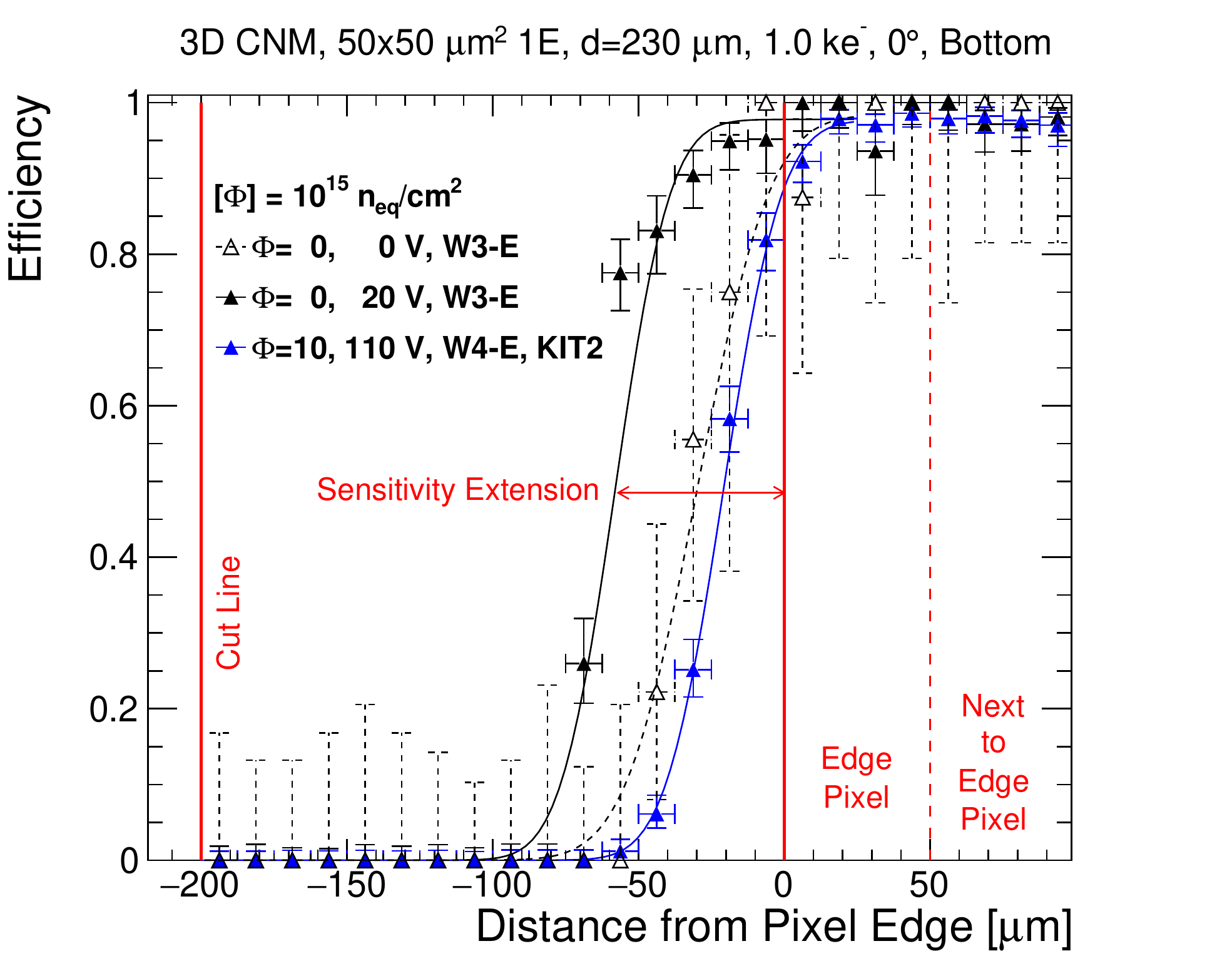}
	\includegraphics[width=7.5cm]{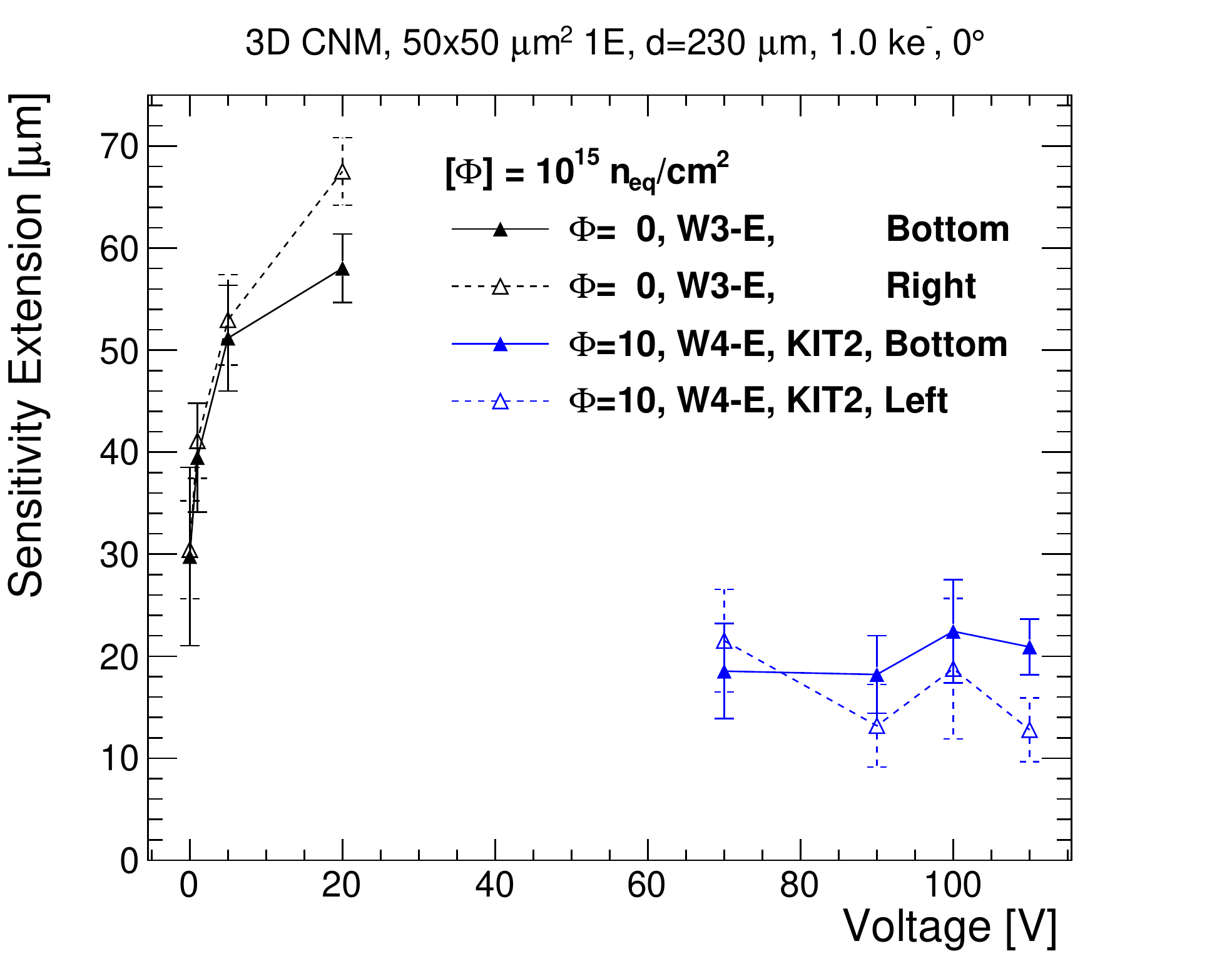}
	\caption{Left: edge efficiency around the last pixel before irradiation at 0 and 20\,V and at a fluence of 1.0$\times10^{16}$\,n$_{\mathrm{eq}}$/cm$^2$ at 110\,V, at 0$^{\circ}$ tilt and 1.0\,ke$^{-}$ threshold. Right: the sensitivity extension of the efficiency beyond the edge of the last pixel as a function of voltage. Uncertainties shown are statistical.}
	\label{fig:edgeEfficiency}
\end{figure}

For devices with the edge termination design without a 3D guard ring (W3-E before irradiation and W4-E irradiated uniformly to 1.0$\times10^{16}$\,n$_{\mathrm{eq}}$/cm$^2$ at KIT), dedicated studies of the edge efficiency were undertaken. A corner of the devices (bottom left or bottom right with respect to the orientation shown in figure~\ref{fig:3Dgeometry} bottom) was placed in overlap with the telescope acceptance. As mentioned in section~\ref{sec:generalPerformance}, the edge pixels of the 3D FE-I4 prototypes were found noisier than in the central region, so that many of them had to be masked leading to a reduced amount of data and large statistical uncertainties.

Figure~\ref{fig:edgeEfficiency} left shows the bottom-edge efficiencies evaluated in the central ROI starting from the cut line 200\,$\mu$m away from the border of the edge pixel up to the next-to-edge pixel. It can be seen that the sensitive region extends beyond the border of the edge pixel. This sensitivity extension, evaluated at an efficiency of 0.5 using Gaussian error function fits, is shown in figure~\ref{fig:edgeEfficiency} right as a function of the bias voltage for the different devices at the bottom and left or right edges. The different edges of the same device (bottom vs. left/right) agree well with each other as expected due to the symmetry. For the unirradiated device, the sensitivity extension amounts to 30\,$\mu$m already at 0\,V, increasing up to about 60\,$\mu$m at 20\,V. After 1.0$\times10^{16}$\,n$_{\mathrm{eq}}$/cm$^2$, it is reduced to about 20\,$\mu$m independent of voltage in the range studied between 70 and 110\,V. This means that the insensitive edge region of these devices is less than 200\,$\mu$m, which is important for abutting the sensors close to each other with small insensitive area between them in the final pixel detector. For devices with guard ring ("C"), the edge efficiency was not studied here, but from previous results it is expected that the sensitive region ends at the border of the edge pixel due to the guard ring~\cite{bib:AFP3D2}, leading to 200\,$\mu$m insensitive edge. In the future, it will be tried to cut the sensors closer to the edge pixels to further reduce the dead region, but attention has to be drawn to the impact on leakage current and yield.


\section{Power dissipation}
\label{sec:Power}

\begin{figure}[hbtp]
	\centering
	\includegraphics[width=7.5cm]{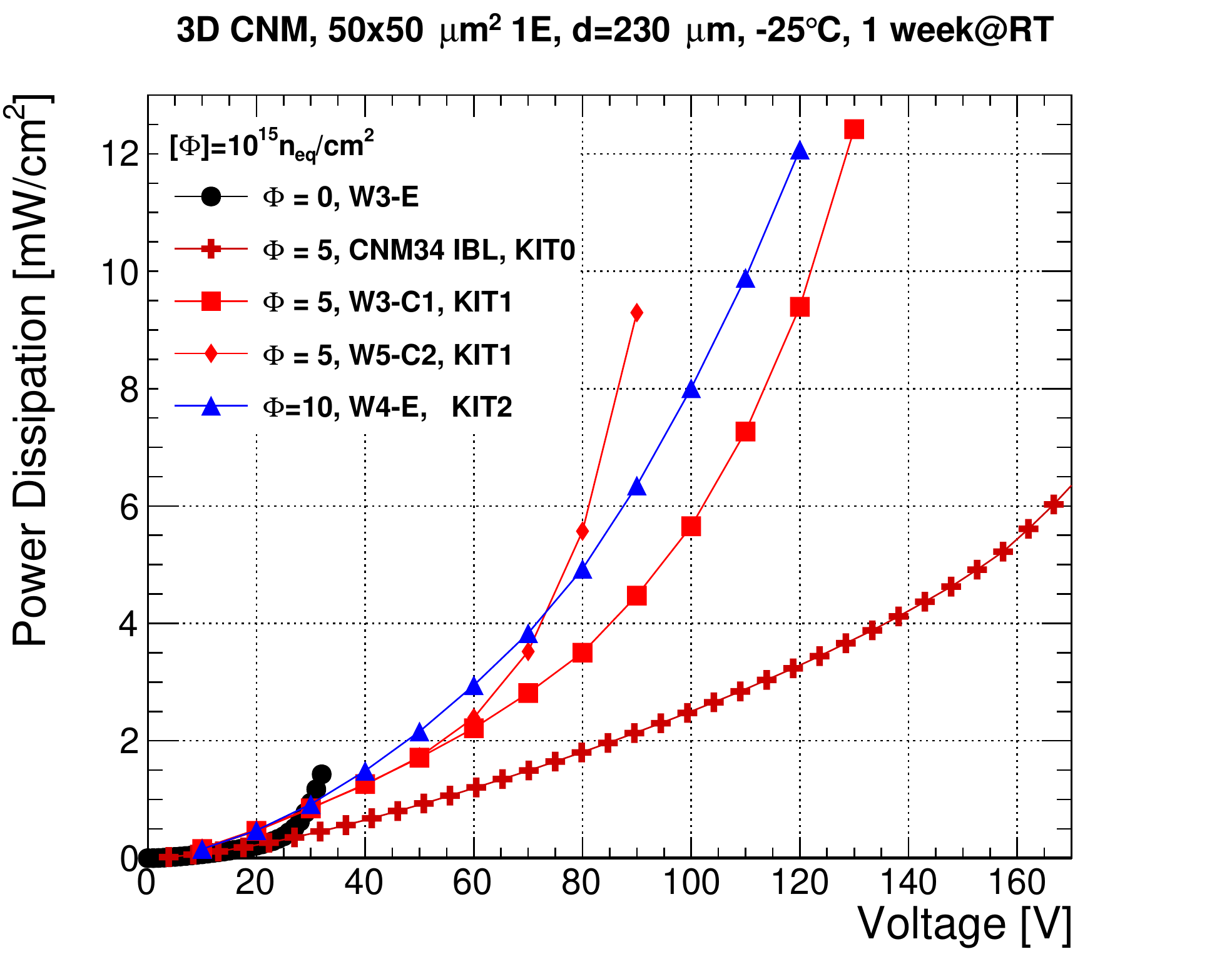}
	\includegraphics[width=7.5cm]{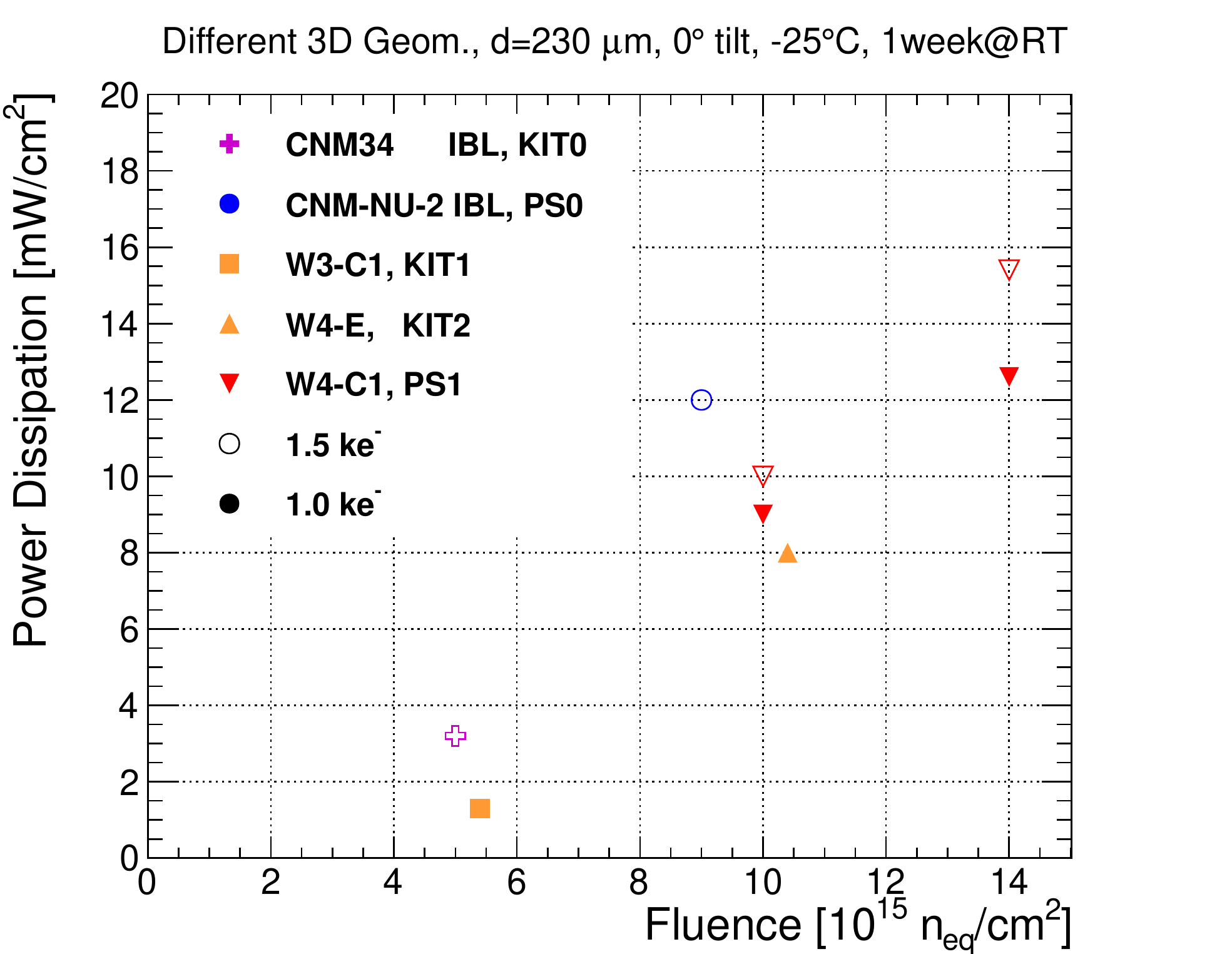}
	\caption{Left: power dissipation as a function of voltage for uniformly irradiated devices. Right: power dissipation evaluated at the operation voltage estimator V$_{97\%}$ as a function of fluence for different devices. The small-pitch 50$\times$50\,$\mu$m$^{2}$ 1E devices (red/orange) are compared to the IBL 50$\times$250\,$\mu$m$^{2}$ 2E generation (blue/magenta). Open markers refer to a threshold of 1.5\,ke$^{-}$, full ones to 1.0\,ke$^{-}$.}
	\label{fig:Power}
\end{figure}

The sensor power dissipation (product of leakage current and bias voltage) is an important parameter to be taken into account for the design of the cooling system and thermal properties of the pixel modules and staves. In fact, given a certain system with fixed thermal properties, the sensor can heat up if the sensor power is too high, which causes even higher power and heating and can hence end up in a so-called thermal runaway. Thus, the power should be as low as possible and must not exceed certain limits given by the system. 

From one pixel device, the power density for a given fluence can be only evaluated properly for uniformly irradiated devices (KIT).
For the non-uniformly irradiated devices (PS), however, V$_{97\%}$ at a certain fluence was combined with the leakage current of uniformly neutron irradiated 3D strip test structures with the same 3D unit cell size~\cite{bib:CNMsmallPitch1} at the same fluence to obtain an estimate of the power dissipation at the operation voltage.
 
Figure~\ref{fig:Power} left shows the power density as a function of voltage for the devices uniformly irradiated at KIT as obtained from the leakage current (section~\ref{sec:leakageCurrent}) at -25$^{\circ}$C after one week annealing at room temperature. The results evaluated at V$_{97\%}$ are presented as a function of fluence in figure~\ref{fig:Power} right, together with the estimates obtained from the pixel devices non-uniformly irradiated at PS in combination with the strips. At 1.0$\times10^{16}$\,n$_{\mathrm{eq}}$/cm$^2$, the results of these two methods agree reasonably well. For the new small-pitch pixel devices at 1.0\,ke$^{-}$ threshold, the power density increases from 1.5\,mW/cm$^{2}$ at 5$\times10^{15}$\,n$_{\mathrm{eq}}$/cm$^2$ over 8--10\,mW/cm$^{2}$ at 1.0$\times10^{16}$\,n$_{\mathrm{eq}}$/cm$^2$ to 13\,mW/cm$^{2}$ at 1.4$\times10^{16}$\,n$_{\mathrm{eq}}$/cm$^2$. As expected, due to the lower operation voltage V$_{97\%}$, this is lower than for the IBL 3D generation, as shown, and much lower than for planar pixel sensors, which were found to dissipate at least 25\,mW/cm$^{2}$ at 1.0$\times10^{16}$\,n$_{\mathrm{eq}}$/cm$^2$~\cite{bib:planarPower}.


\section{Conclusions and outlook}
\label{sec:conclusions}

A new generation of radiation-hard 3D pixel detectors with small pixel sizes of 50$\times$50 and 25$\times$100\,$\mu$m$^{2}$ and hence reduced inter-electrode distances is being developed in the framework of the HL-LHC pixel detector upgrades. For the first time, silicon pixel detectors have been studied and shown a good performance up to very high particle fluences of $3\times10^{16}$\,n$_{\mathrm{eq}}$/cm$^2$, beyond the full expected ATLAS ITk fluence and more than double the baseline scenario foreseeing one replacement. 

These first tests were performed on small-pitch 3D sensor prototypes of 230\,$\mu$m thickness produced by CNM Barcelona, which were connected to the FE-I4 readout chip with larger pixel size. Irradiation campaigns with protons were carried out at KIT and CERN-PS and the devices responsive after irradiation have been measured in beam tests at CERN-SPS.

Despite being specified only to IBL fluences, the FE-I4 readout chip survived in most of the cases and worked reasonably well after such high irradiation for devices with 50$\times$50\,$\mu$m$^{2}$ 1E sensor pixels. A fraction of pixels had to be excluded from data taking and analysis. The charge collection efficiency, as obtained from the time-over-threshold, was found to be high. Before irradiation, 98\% hit efficiency was found already for an unbiased sensor. Even after the highest fluence studied of $2.8\times10^{16}$\,n$_{\mathrm{eq}}$/cm$^2$, a 98\% plateau hit efficiency was obtained. The voltage needed to reach the ITk benchmark efficiency of 97\% was determined as about 40, 100 and 150\,V for 0.5, 1.4 and $2.8\times10^{16}$\,n$_{\mathrm{eq}}$/cm$^2$ at a tilt of 0$^{\circ}$ and 1.0\,ke$^{-}$ threshold. This was shown to be significantly lower than for the previous IBL 3D generation with larger inter-electrode distances, and in particular than for thin planar sensors. This also translates into an improvement regarding the power dissipation, where 8--10\,mW/cm$^{2}$ were obtained at $1.0\times10^{16}$\,n$_{\mathrm{eq}}$/cm$^2$ for 3D small-pitch sensors (compared to more than 25\,mW/cm$^{2}$ needed for planar sensors), and 13\,mW/cm$^{2}$ at $1.4\times10^{16}$\,n$_{\mathrm{eq}}$/cm$^2$.
Sensors without 3D guard ring were observed to be sensitive up to 60\,$\mu$m beyond the last pixel before irradiation, resulting in less than 200\,$\mu$m inactive edge.

To conclude, it was shown that small-pitch 3D sensors meet the requirements of the innermost pixel layer of the ATLAS ITk detector and were thus chosen as the baseline sensor technology. Even at fluences of $3\times10^{16}$\,n$_{\mathrm{eq}}$/cm$^2$, significantly beyond the ITk baseline fluence of $1.3\times10^{16}$\,n$_{\mathrm{eq}}$/cm$^2$ (assuming one replacement)~\cite{bib:ITkPixelTDR}, an excellent radiation hardness of 3D sensors was demonstrated, with no limit observed so far, making them also promising potential candidates for applications with even higher fluence requirements such as the LHCb phase II upgrade or future hadron circular colliders.

In the future, tests with these 3D FE-I4 prototypes at even higher fluences are planned. In parallel, studies with small-pitch 3D pixel sensors connected to the analogue ROC4SENS readout chip with 50$\times$50\,$\mu$m$^{2}$ pixel size are on-going~\cite{bib:SantanderRoc4Sense}. First results up to $3\times10^{15}$\,n$_{\mathrm{eq}}$/cm$^2$ are consistent with this study. Moreover, 3D sensor productions compatible with the newly developed RD53A readout chip optimised for applications at the HL-LHC with 50$\times$50 and 25$\times$100\,$\mu$m$^{2}$ pixel size are carried out at CNM Barcelona. An even improved performance is expected due to lower thresholds and improved radiation hardness. Different sensor thicknesses between 75 and 200\,$\mu$m will be studied. 
Also other vendors like Fondazione Bruno Kessler (FBK), Trento, Italy~\cite{bib:FBKsmallPitch}, and SINTEF, Trondheim, Norway~\cite{bib:SINTEF}, have productions and studies on-going for HL-LHC 3D pixel detectors.

\acknowledgments 

The authors wish to thank A.\,Rummler, M.\,Bomben, J.\,Weingarten and the other ATLAS ITk beam test participants for great support and discussions at the beam tests; also to F.\,Ravotti, G.\,Pezzullo and B.\,Gkotse (CERN-PS IRRAD) as well as A.\,Dierlamm and F.\,B\"{o}gelspacher (KIT) for excellent support for the irradiations. This work was partly performed in the framework of the CERN RD50 collaboration.This work was partially funded by: the MINECO, Spanish Government, under grants FPA2013-48308-C2-1-P, FPA2015-69260-C3-2-R, FPA2015-69260-C3-3-R (co-financed with the European Union's FEDER funds) and SEV-2012-0234 (Severo Ochoa excellence programme), under the Juan de la Cierva programme and under the Spanish ICTS Network MICRONANOFABS partially supported by MINECO; the Catalan Government (AGAUR): Grups de Recerca Consolidats (SGR 2014 1177); "la Caixa" INPhINIT Fellowship Grant for Doctoral studies at Spanish Research Centres of Excellence, "la Caixa" Banking Foundation, Barcelona, Spain; and the European Union's Horizon 2020 Research and Innovation programme under Grant Agreement no. 654168 (AIDA2020) and under the Marie Sklodowska-Curie grant agreement no. 713673.



\providecommand{\href}[2]{#2}\begingroup\raggedright\endgroup

\end{document}